\documentclass[a4paper]{ar-1col}

\usepackage{natbib}
\usepackage{aas_macros}
\usepackage{amssymb}
\usepackage{url}
\usepackage{subcaption}
\usepackage{graphics}

\def\bfnabla{{\mbox{\boldmath $\nabla$}}}
\newcommand\bv{{\mbox{\boldmath $v$}}}
\newcommand\bb{{\mbox{\boldmath $B$}}}

\jname{Annu .Rev. Astron. Astrophys.}
\jvol{58}
\jyear{2020}
\doi{10.1146/annurev-astro-081817-051905}

\begin{document}

\markboth{Davis \& Tchekhovskoy}{Disks and Jets in AGNs}

\title{Magnetohydrodynamic Simulations of Active Galactic Nucleus Disks and Jets}

\author{Shane W. Davis$^1$ and Alexander Tchekhovskoy$^2$
\affil{$^1$Department of Astronomy, University of Virginia, Charlottesville, Virginia, 22904, USA; email: swd8g@virginia.edu}
\affil{$^2$Center for Interdisciplinary Exploration and Research in Astrophysics (CIERA), Department of Physics \& Astronomy, Northwestern University, Evanston, Illinois, 60657; email: atchekho@northwestern.edu}}

\begin{abstract}
  There is a broad consensus that accretion onto supermassive black holes and consequent jet formation power the observed emission from active galactic nuclei (AGNs). However, there has been less agreement about how jets form in accretion flows, their possible relationship to black hole spin, and how they interact with the surrounding medium. There have also been theoretical concerns about instabilities in standard accretion disk models and lingering discrepancies with observational constraints.  Despite seemingly successful applications to X-ray binaries, the standard accretion disk model faces a growing list of observational constraints that challenge its application to AGNs. Theoretical exploration of these questions has become increasingly reliant on numerical simulations owing to the dynamic nature of these flows and the complex interplay between hydrodynamics, magnetic fields, radiation transfer, and curved spacetime. We conclude the following:
\begin{minipage}{3.93in}
\vspace{0.1in}
\begin{itemize}
\item
The advent of general relativistic magnetohydrodynamics (MHD) simulations has greatly improved our understanding of jet production and its dependence on black hole spin.\\

\item
 Simulation results show both disks and jets are sensitive to the magnetic flux threading the accretion flow as well as possible misaglingment between the angular momentum of the accretion flow and the black hole spin.\\

\item
Radiation MHD simulations are providing new insights into the stability of luminous accretion flows and highlighting the potential importance of radiation viscosity, UV opacity from atoms, and spiral density waves in AGNs.\\
\end{itemize}
\end{minipage}
\end{abstract}

\begin{keywords}
black hole physics, radiation transport, general relativity
\end{keywords}
\maketitle

\tableofcontents

\section{INTRODUCTION}

The early observations of quasars and the realization that they must be distant objects \citep{1963Natur.197.1040S} already indicated a need for objects much smaller than galaxies to be producing immense energies to power their optical emission and radio lobes. \begin{marginnote}[-2pt]\entry{Quasar}{a luminous active galactic nucleus that outshines its host galaxy}\end{marginnote}  Accretion of interstellar gas onto a supermassive black hole was identified as the only viable sources for such energy \citep{1964ApJ...140..796S,1969Natur.223..690L}.

The first detailed model of the vertical and radial structure of disks with angular momentum transport was provided by \citet{1973A&A....24..337S} with additional insights from \citet{1974MNRAS.168..603L} and general relativistic generalization by \citet{1973blho.conf..343N}.  The key assumptions in the model were that the  vertical scale height of the disk was small compared with the radius everywhere and that the stress leading to angular moment transport could be approximated as a constant fraction ($\alpha$) of the total pressure in the disk midplane. Even 45 years later, this seminal work remains the primary basis of our understanding of accretion onto black holes in XRBs\begin{marginnote}[-2pt]\entry{X-ray Binary (XRB)}{X-ray emitting source powered by accretion from companion star onto a black hole or neutron star}\end{marginnote}and AGN\begin{marginnote}[-2pt]\entry{Active galactic nucleus (AGN)}{compact emitting region in the center of a galaxy thought to be powered by accretion onto a supermassive black hole}\end{marginnote}and we refer to this as the standard model of accretion disks.  Although these steady disk models suffer from instabilities discussed below, they were successful in broadly explaining the spectral properties of XRBs \citep[see e.g.][]{2006ARA&A..44...49R,2007A&ARv..15....1D} and the optical/UV continuum of luminous AGNs \citep{1978Natur.272..706S,1984AdSpR...3..249K}.

Although this standard model is thought to apply in the luminous accretion regime, there are observational and theoretical arguments that this model breaks down when the accretion rate is very low or high when compared with the Eddington (or critical accretion) rate,
\begin{equation}
\dot{M}_{\rm E} = \frac{4 \pi G M m_p}{\eta \sigma_{\rm T} c},
\end{equation}
where $M$ is the mass of the black hole, $\eta$ is the radiative efficiency, and other values are fundamental constants with standard meanings.  This is the accretion rate that would produce an Eddington luminosity\begin{marginnote}[-2pt]\entry{Eddington luminosity}{The luminosity for which outward radiation pressure on the electrons equals the inward gravitational pull on the ions}\end{marginnote}of emission ($L_{\rm E} \simeq 10^{38} M/M_{\odot}$) if the radiative efficiency remains constant (typically assumed to be around 10\%).  At low accretion rates ($\dot{M} \lesssim 0.01 \dot{M}_{\rm E}$), flows are thought to be hotter and less radiatively efficient \citep{2014ARA&A..52..529Y}.  In these models, number densities are low enough that Coulomb collisions are less efficient at coupling electrons to ions and radiative cooling is weak, leading to flows dominated by radial advection of thermal energy \citep{1994ApJ...428L..13N} and/or outflows \citep{1999MNRAS.303L...1B}.
At high accretion rates ($\dot{M} \gtrsim \dot{M}_{\rm E}$), the assumption of a thin disk is no longer self-consistent as the vertical scale height of the inner disk becomes comparable with radius.  The flow is thought to again become less radiatively efficient as the photon diffusion time becomes longer than the radial inflow time, allowing a larger fraction of the dissipated energy to be advected into the black hole.  Models including these effects are called slim disks \citep{1988ApJ...332..646A}.

All these accretion disk models rely on an anomalous viscosity or stress ($\alpha$ prescription) to explain angular momentum transport because realistic physical viscosities are too low by many orders of magnitude. Much of the theoretical work in the field for several decades focused on discerning the origin of this anomalous stress.\begin{marginnote}[-2pt]\entry{Magnetorotational instability (MRI)}{ Instability driven by magnetic fields interacting with shear that provides a source of angular momentum transport}\end{marginnote}After the identification of the role of the \citep[MRI;][]{1991ApJ...376..214B,1998RvMP...70....1B} in driving magnetohydrodynamic (MHD) turbulence in disks, it is now widely believed that MRI turbulent stresses provide the dominant angular momentum transport mechanism.\begin{marginnote}[-2pt]\entry{Magnetohydrodynamics (MHD)}{The combined magnetic and fluid dynamics of conducting plasmas}\end{marginnote} To date, this has not led to a significant revision of the standard model because the Maxwell and Reynolds stresses provided by the MRI turbulence are broadly consistent with the $\alpha$ prescription ansatz \citep{1999ApJ...521..650B}. However, in addition to disk internal stresses, disk outflows can also remove the angular momentum from the accretion disk and affect the disk structure. This is particularly true in the case of MHD-driven outflows, where nonlocal stresses are dominant \citep{1982MNRAS.199..883B}. This suggests that the processes of accretion and outflows may be tightly coupled.

Observations indicate that accretion systems produce various types of outflows. Of particular interest are collimated outflows, more commonly referred to as jets. \begin{marginnote}[-2pt]\entry{Blandford-Znajek (BZ)}{Mechanism by which magnetic fields are twisted by black hole spin to form an outflow}\end{marginnote} Assisted with large-scale magnetic fields, jets can derive their power from the rotation of either the black hole via the \citet{1977MNRAS.179..433B} effect or the rotation of the accretion disk via the \citet{1982MNRAS.199..883B} mechanism. The jets can cover many decades in radius, spanning from the event horizon of the black hole to the outskirts of the galaxies and beyond. Modeling jets is challenging owing to their nonlinear nature that gets exacerbated by the poorly understood disk-jet connection near their base and turbulent interactions with the ambient medium as they propagate away from the center.

Although work on the revision and development of analytic models continues, the dynamic and three-dimensional nature of the problem of coupled radiation transfer and MHD in general relativistic spacetimes has proven to be a challenge for further analytic progress and simpler modeling efforts. The critical role played by dynamic magnetic fields in the launching of jets and evolution of accretion disks has led to an increased focus and reliance on numerical simulation modeling.  In this review, we summarize recent developments and the current status of numerical simulations of disks and jets.  We review observational results that challenge the existing models and motivate the incorporation of new or improved treatments of physical processes in state-of-the-art simulations.

\section{CURRENT STATUS OF OBSERVATION AND THEORY}

\subsection{Accretion Disk Theory in AGNs}
\label{theorystatus}

There are numerous presentations of the standard disk model equations, so we do not attempt to reproduce them here, but in our opinion the seminal paper by \cite{1973A&A....24..337S} remains the best place to start.  For a detailed discussion of how this model applies to AGNs, we recommend the textbook by \citet{1999agnc.book.....K}.  Here, we briefly review the key results relevant to this work.  Broadly speaking, the accretion flow divides into two regions: an inner region, in which radiation pressure is the primary support against gravity, and an outer region, in which gas pressure dominates.  The transition radius is a function of disk parameters and is given by
\begin{equation}
r_{\rm tr} \simeq 340 \; r_g 
\left(\frac{0.1}{\eta}\;\frac{L}{0.1L_{\rm
      E}}\right)^{16/21}\left( \frac{\alpha}{0.1} \; \frac{M}{10^8 M_\odot}\right)^{2/21}
\;\frac{R_R^{6/7}}{R_z^{10/21}R_T^{2/21}}\bigg|_{r_{\rm tr}}, \label{eq:rtransition}
\end{equation}
where $r_g = GM/c^2$ is the gravitational radius, and $\alpha$ is the famous parameter that relates the accretion stress $\tau_{r\phi}$ to the total midplane pressure via $\tau_{r\phi} = \alpha P_{\rm tot}$.  The radial- and black-hole-spin-dependent quantities $R_R$, $R_z$, and $R_T$ are defined in chapter 7 of \citet{1999agnc.book.....K} and represent general relativistic correction factors.  Here, $P_{\rm tot}$ is the sum of gas pressure $P_{\rm g}$ and radiation pressure $P_{\rm r}$ because the standard model does not consider the impact of magnetic pressure.  One can derive different scalings of quantities of interest in these inner and outer regions.  For brevity, we focus on the scalings in the radiation-pressure dominated inner regions and assume electron scattering dominates the opacity.  We find
\begin{eqnarray}
  T_{\rm eff} & = & 6.2 \times 10^5\; {\rm K} \; \left(\frac{10^8 M_\odot}{M}\;\frac{0.1}{\eta}\;
  \frac{L}{0.1L_{\rm E}}\right)^{1/4} R_R^{1/4} \left(\frac{r}{r_g}\right)^{-3/4}\label{eq:teff}\\
  T_{\rm mid} & = & 9 \times 10^5\; {\rm K} \;  \left( \frac{10^8 M_\odot}{M}\; \frac{0.1}{\alpha}\right)^{1/4} \left(\frac{R_z R_T}{R_R}\right)^{1/4} \left(\frac{r}{r_g}\right)^{-3/8}\label{eq:tmid}\\
  \rho_{\rm mid} & = & 3.2 \times 10^{-13} \; {\rm g/cm^3} \;  \frac{10^8 M_\odot}{M}\;\frac{0.1}{\alpha}\;\left(\frac{\eta}{0.1}\;
  \frac{0.1L_{\rm E}}{L}\right)^2 \frac{R_z^2 R_T}{R_R^3} \left(\frac{r}{r_g}\right)^{3/2},\label{eq:density}
\end{eqnarray}
where $r$ is radius. The effective temperature $T_{\rm eff}$ is an estimate of the surface temperature and $\rho_{\rm mid}$ and $T_{\rm mid}$ are the midplane density and temperature, respectively. The ratio of $h/r$, where $h$ is the disk scale height for the radiation dominated flow is
\begin{equation}
  \frac{h}{r} = 0.75 \;\frac{0.1}{\eta}\;
  \frac{L}{0.1L_{\rm E}}\frac{R_R}{R_z}\frac{r_g}{r}.\label{eq:height}
\end{equation}
(See the sidebar titled Thermal Instability in the Standard Accretion Disk Model.)

\begin{textbox}[ht]
\section{Thermal Instability in the Standard Accretion Disk Model}
The thermal instability of radiation pressure-dominated accretion disks \citep{1976MNRAS.175..613S} can be understood via a simple argument about how the heating and cooling rates in an accretion flow depend on temperature \citep{1978ApJ...221..652P}.  In an accretion flow, the cooling rate per unit mass $Q_-$ from a disk annulus at radius $r$ is simply given by the local radiative flux $F_r$ divided by the mass surface density (mass per unit area) $\Sigma$ so that (accounting for both sides of the disk)
\begin{displaymath}
Q_- = \frac{2 F_r}{\Sigma}.
\end{displaymath}
The rate at which energy is dissipated in a shear flow is related to the product of the accretion stresses $\tau_{r\phi}$ and the shear $r d\Omega/dr$. Integrating over height and dividing by $\Sigma$ yields a heating rate per unit mass 
\begin{displaymath}
Q_+ =  \frac{1}{\Sigma}\int \tau_{r\phi} \left( -r \frac{d\Omega}{dr}\right) dz
\simeq \frac{3 h\tau_{r \phi} \Omega}{\Sigma}.
\end{displaymath}
Using the standard model equations in the radiation pressure-dominated regime with electron scattering opacity yields $h \propto F_r \propto P_r/\Sigma$ and $\tau_{r\phi} \propto P_r$.  Since $P_r =a T^4/3$ we have:
\begin{displaymath}
Q_+ \propto \frac{T^8}{\Sigma^2}, \;
Q_- \propto \frac{T^4}{\Sigma^2}.
\end{displaymath}
In deriving the standard model, we assume the disk is in thermal equilibrium with $Q_-=Q_+$. Now suppose that the temperature is perturbed upward. Because departures from thermal equilibrium happen on the thermal timescale while changes in the surface density happen on the much longer viscous timescale we assume $\Sigma$ is constant.  An increase in the temperature causes a large increase in the heating rate due to the $T^8$ dependence and a somewhat more modest change in the cooling rate due to the $T^4$ dependence.  So, as the temperature is perturbed upward, the heating rate goes up faster than the cooling rate, leading to a thermal runaway.  The same argument applies in reverse to downward perturbations in temperature, and we conclude that radiation pressure-dominated disks are thermally unstable.
\end{textbox}

Although the standard accretion model approximately succeeds at explaining SEDs\begin{marginnote}[-2pt]\entry{SED}{Spectral energy distribution}\end{marginnote} of AGN, there are a number of observational and theoretical issues that have arisen.  Shortly after the model was formulated, it was realized that the model is subject to inflow (or viscous) and thermal instabilities \citep{1974ApJ...187L...1L,1976MNRAS.175..613S} when radiation pressure exceeds gas pressure in the disk midplane. Inserting the above scalings into standard expression for the radiation and gas pressure yields the scaling,
\begin{equation}
\frac{P_r}{P_g} \propto \alpha^{1/4}M^{1/4}\eta^{-2}\left(\frac{L}{L_{\rm
      E}}\right)^{2} \left(\frac{r}{r_g}\right)^{-21/8}.
\end{equation}
Hence, for the same Eddington ratio, the radiation pressure in at $10^9 M_\odot$ black hole is expected to 100 times more dominated by radiation pressure than a black hole with $10 M_\odot$.  This extreme dominance of radiation pressure in AGNs could suggest that the thermal and inflow instabilities could have a larger impact in AGNs. It is generally thought that such instabilities will lead to limit cycle behavior in which the disk oscillates between a gas dominated lower branch and an advection dominated upper branch \citep{2000ApJ...535..798N,2002ApJ...576..908J}.  The relevant timescales are the dynamical time, thermal time, and inflow (viscous) time given by
\begin{eqnarray}
  t_{\rm dyn} & = & \frac{1}{\Omega} =  1 \; {\rm day} \frac{M}{10^8 M_\odot} \left(\frac{r}{30 r_g}
  \right)^{3/2}\;\label{tdynamical}\\
  t_{\rm th} & = & \frac{1}{\alpha \Omega} = 9 \; {\rm day} \frac{M}{10^8 M_\odot}\frac{0.1}{\alpha}
  \left(\frac{r}{30 r_g}\right)^{3/2}\label{eq:tthermal}\\
  t_{\rm in} & = & \frac{1}{\alpha \Omega}\frac{r^2}{h^2} =  260 \; {\rm yr} \frac{M}{10^8 M_\odot}
  \frac{0.1}{\alpha}\left(\frac{h}{0.01 r}\right)^2\left(\frac{r}{30 r_g}\right)^{3/2},\label{eq:tinflow}
\end{eqnarray}
where $\Omega=(GM/r^3)^{1/2}$ is the Keplerian rotation rate and $h/r$ is the aspect ratio of the disk scale height to radius. Time-dependent models yield evolution along stable branches occurring on the inflow timescale, whereas evolution between branches occurs on the thermal timescale.  If the disk is as thin as commonly thought, the inflow timescale is very long compared to observing timescales for most systems.

It was shown that the thermal and inflow instabilities could be suppressed in alternative stress prescriptions \citep{1981ApJ...247...19S} where the stress scaled not with the total pressure but gas pressure only.  However, as we discuss below, evidence from radiation MHD simulations of accretion flows suggests the stress does generally scale with the total pressure and, thus, provide some evidence that thermal instability is present \citep{2013ApJ...778...65J,2016MNRAS.463.3437M}. 

\subsection{Observational Status: AGNs Versus XRBs}

\begin{figure}[h]
\vskip -144pt
\includegraphics[width=5in]{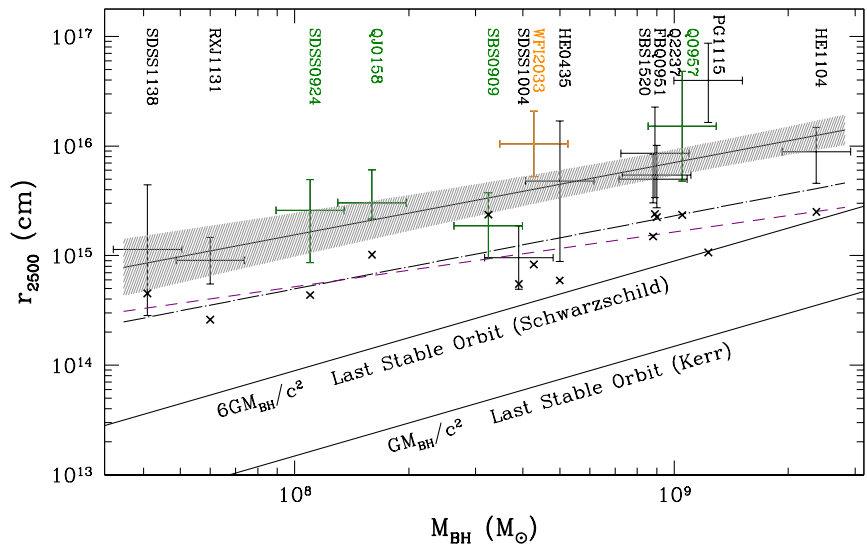}
\vskip -90pt

\caption{Discrepancies between microlensing measured sizes and predictions from the standard model.  The solid line gives the best fit relation and the shaded regions shows the $1\sigma$ uncertainties.  For comparison, the black dot-dashed curve provides the estimate for thin accretion disk models with the same mass (assuming $L/L_{\rm E}=1/3$, $\eta=0.1$).  The observed relation is about a factor of 3-4 larger than the standard model prediction. For comparison, the solid lines labeled "Last stable orbit" provide the size of the innermost stable circular orbit for non-spinning and maximally spinning black holes as a function of mass. Figure adapted from \citet{2018ApJ...869..106M} with permission.} \label{f:sizes}
\end{figure}

On the observational side, there are also a number of potential discrepancies. \citet{1999PASP..111....1K} provide a somewhat dated but still useful review of several of the problems, many of which remain unresolved.  Some of the issues that have received the most attention are the apparent discrepancies between the standard model and sizes of emission regions determined by microlensing \citep[e.g][]{2010ApJ...712.1129M,2011ApJ...729...34B,2018ApJ...869..106M} and continuum reverberation mapping \citep[see e.g.][]{2015ApJ...806..129E,2017ApJ...836..186J,2019ApJ...880..126H}.  A summary of the microlensing size scale constraints from \citet{2018ApJ...869..106M} is shown in Figure~\ref{f:sizes}.  Their best-fit relation suggests that emitting regions are about a factor of 3-4 greater than the standard model predicts, which is consistent with the time delays inferred from continuum reverberation mapping.

Variability is also a challenge for disk models. Although the XRB GRS 1915+109 famously shows variability that looks somewhat like the limit cycle predictions \citep{2000ApJ...535..798N}, XRBs show no generic evidence of such variability \citep{2007A&ARv..15....1D}. In AGNs, the predicted viscous timescales are expected to be too long (in some cases, thousands of years) to probe observationally.  Instead, the observed variability poses a different challenge. Surveys and monitoring campaigns have identified a large number of ``changing look'' AGNs, that show  large-amplitude variations in the continuum or emission lines on timescale of months to years \citep{2015ApJ...800..144L,2016MNRAS.457..389M}, which are difficult to explain with ``viscous'' evolution timescales \citep{2018NatAs...2..102L}.  The question becomes how does one obtain large-amplitude variability of accretion disks on such short timescales?

\begin{figure}[h]
  \begin{subfigure}{.5\textwidth}
    \centering
    \includegraphics[width=2.5in]{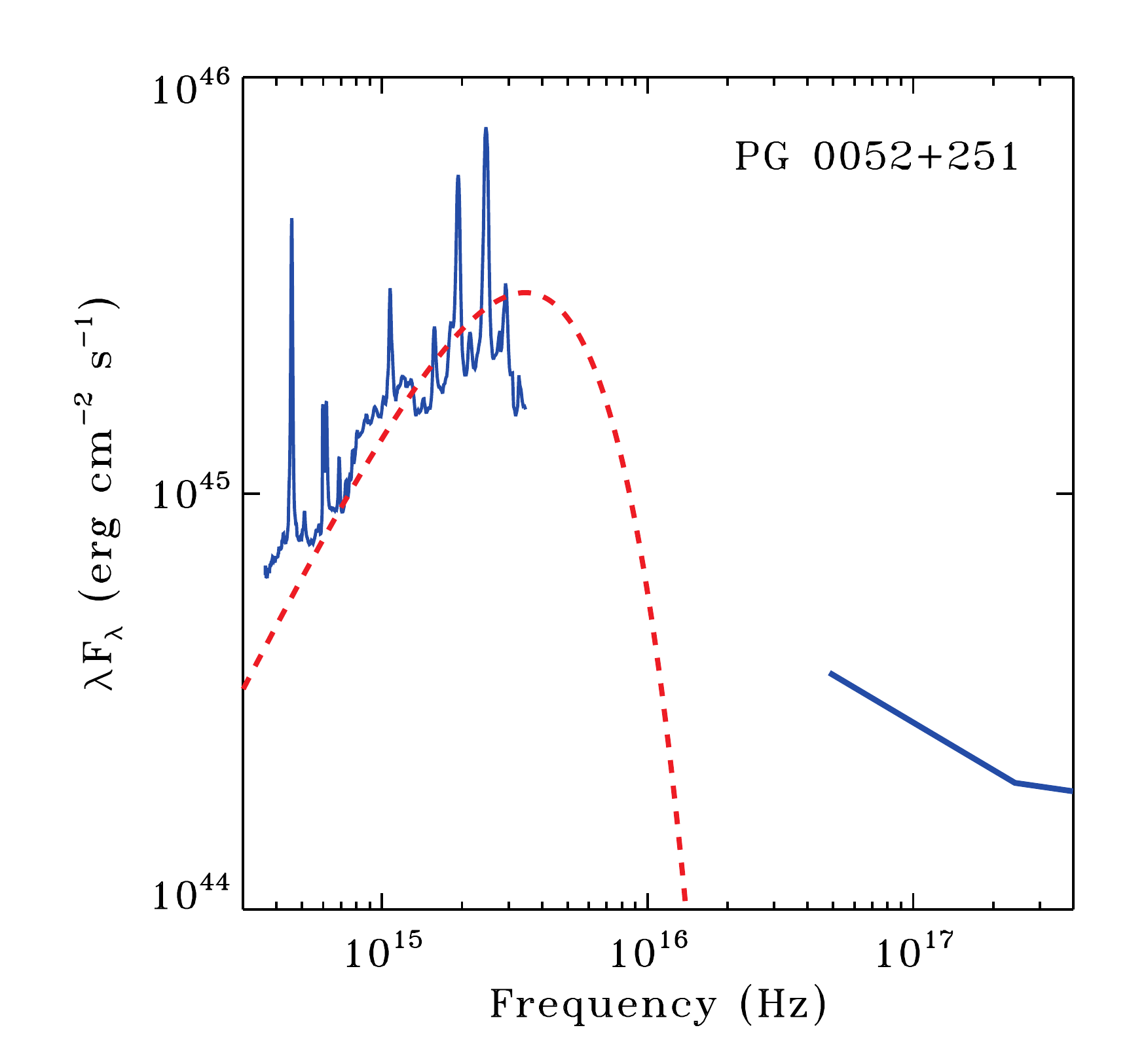}
  \end{subfigure}
  \begin{subfigure}{.5\textwidth}
    \centering
    \includegraphics[width=2.3in]{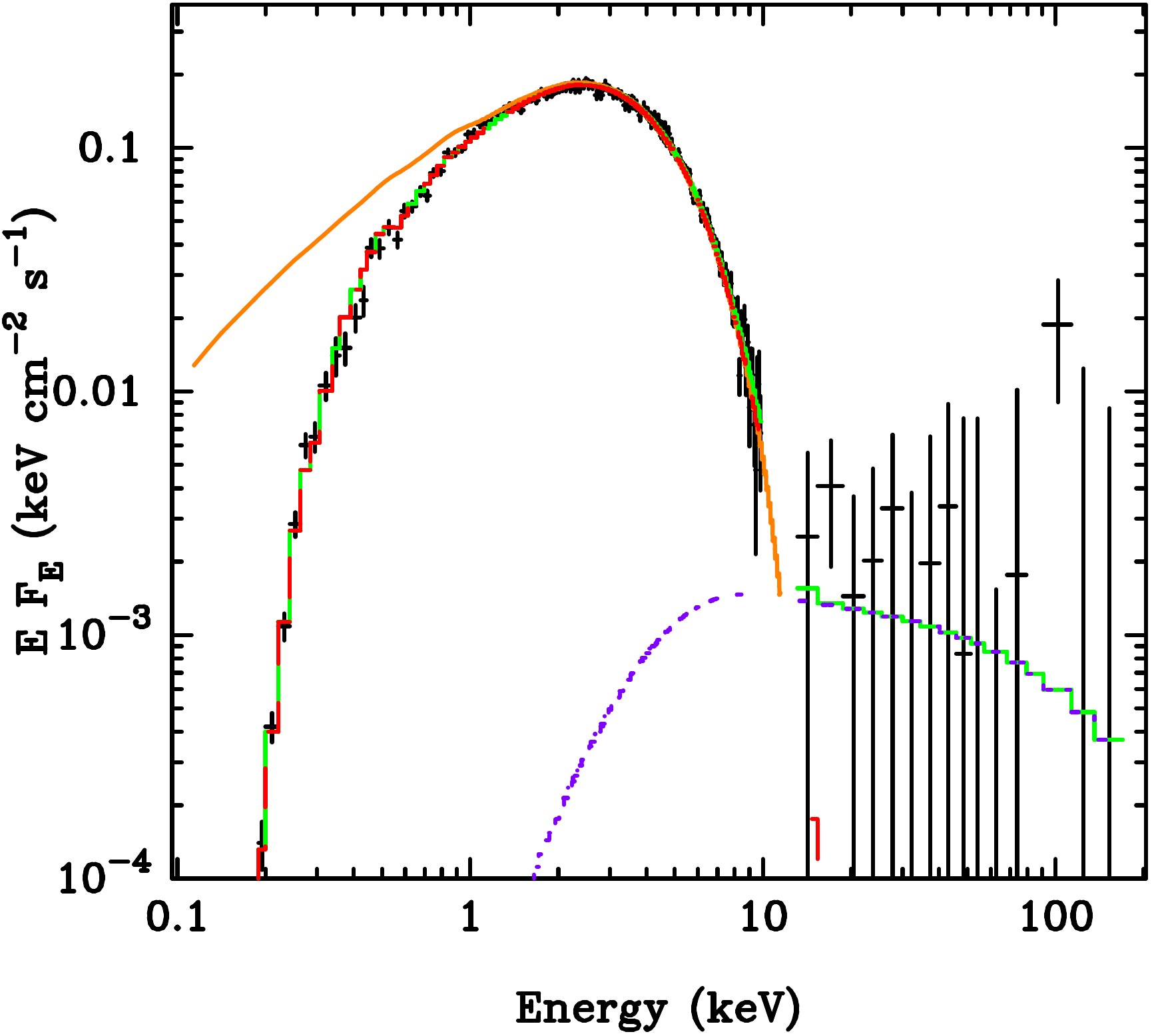}
  \end{subfigure}
\caption{{\it Left panel:} Comparison of observed broadband quasar spectral energy distributions from PG 0052+251 \citep{2005ApJ...619...41S} with relativistic multitemperature blackbody models chosen to match the reverberation-mapped mass and with accretion rates chosen to match the optical emission. The blue X-ray curves show the level and slope change of the X-ray emission. Note that there is a turnover in the data in the ultraviolet band around $3 \times 10^{15}$Hz (1000\AA), whereas the model (red dashed curve)  continues rising into the extreme ultraviolet. This is characteristic of most of the quasars in the PG sample \citep{2011ApJ...728...98D}. {\it Right panel:} The best-fit model to the black hole X-ray binary source LMC X-3 \citep{2006ApJ...647..525D}.  The {\it BeppoSAX} data are shown in black, whereas the best-fit model is shown as a green curve.  The model has two components, with the best-fit absorbed relativistic accretion disk model shown in red.  The unabsorbed relativistic accretion disk model is show in orange. The model provides a good fit to the data with only two free parameters in the disk model.}
\label{f:spectra}
\end{figure}

Others have looked at the spectral variations of the UV continuum of AGNs with mass estimates and found them to be poorly fit by spectral models based on the $\alpha$-disks \citep[see e.g.,][]{1984AdSpR...3..249K,1989MNRAS.238..897L,1997ApJ...476..605S,2007ApJ...659..211B,2007ApJ...668..682D}.  The discrepancy arises in the sense that the model spectra rise too rapidly into the far- to extreme-UV, whereas the observed spectra tend to be flat or declining in $\lambda F_\lambda$ above $\sim 1000$\AA.  As a typical example, PG 0052+251 is shown in the left panel of Figure~\ref{f:spectra}. Here, the comparison relativistic blackbody disk model has been chosen to have the best-fit reverberation mapping mass and an accretion rate scaled to match the spectrum in the optical band.   This has led to the consideration that some other physics is at play, with speculations including advection \citep{2014MNRAS.438..672N}, reprocessing close to the inner disk \citep{2012MNRAS.423..451L}, or outflows \citep{2012MNRAS.426..656S,2014MNRAS.438.3024L}.  However, this is also the wavelength range in which dust reddening in the AGN host galaxy might be important, and much of the discrepancy might be accounted for with dust reddening \citep[e.g.,][]{2015MNRAS.446.3427C}.  Unfortunately, the precise form of the reddening curve and the amount of extinction remain subjects of debate \citep{2003AJ....126.1131R,2004ApJ...616..147G,2016ApJ...832....8B}.  Other significant problems include the lack of evidence for Lyman edge features \citep{2012ApJ...752..162S}, which tends to be prominent (in either emission or absorption depending on model parameters) in detailed spectral models \citep{2001ApJ...559..680H}.

The overall picture is one in which the standard accretion disk model roughly predicts the presence of the optical to UV continuum but seems at odds with a number of observational constraints when considered in detail.  This situation contrasts with the case of XRBs, which the standard model was developed to explain \citep{1973A&A....24..337S}.  The phenomenology of black hole XRBs can be rather complicated, but there are a number of good reviews \citep[see e.g.,][]{2006ARA&A..44...49R,2007A&ARv..15....1D}. The spectral evolution is usually described in terms of a number of spectral states, with the high/soft (thermal dominant) and low/hard states being the best-characterized states based on the differences in their spectral shape, count rates, and variability.  The high/soft state features a prominent broad thermal continuum component, which is generally assumed to be emission from a geometrically thin, optically thick accretion disk, whereas the low/hard states is dominated by harder X-ray continuum component that is generally thought to be produced by inverse Compton scattering of photons in a hot corona above or interior to the disk.  Although XRBs can have significant variability on shorter timescales, this is mostly associated with the coronal component.  The high/soft state tends to be characterized by relatively low variability, particularly in disk-like component of emission.

If one focuses solely on the seemingly disk-dominated high/soft state, then the predictions of the standard model seem to match the data well \citep{2005ApJ...621..372D,2011MNRAS.411..337D}.  A good example of this, LMC X-3, is shown in the right panel of Figure~\ref{f:spectra} with the best-fit model being a fully relativistic spectrum derived from self-consistent radiation transfer calculations through a vertically varying disk structure \citep{2005ApJ...621..372D}.  Therefore, a key question is why is there this seeming discrepancy in how well standard disk models reproduce the properties of AGNs and black hole XRBs?  One possibility is that we have overstated the agreement between the models and XRB observations. Although there are compelling models for the origin of the coronae and the nature of the transitions between the different spectral states  \citep[e.g.,][]{1997ApJ...489..865E,2009PASJ...61L...7O}, there is no consensus and this remains an active area of research.  Whether or not one accepts the premise that standard models work for XRBs and fail for AGNs, there are several reasons to expect that accretion flows in AGNs and XRBs might differ.  We focus on three possibilities.

The first possibility is that the standard model predicts that disks around supermassive black holes should be more radiation pressure-dominated than those around stellar mass black holes. For accretion rates typically inferred in high/soft state XRBs, these flows are expected to be radiation pressure-dominated, but possibly only by factors of $\sim 10$ or less.  In contrast, the inner radii of the most massive and luminous AGNs could be radiation-dominated by factors of a million or more.  It is therefore conceivable that the impact of thermal and inflow instabilities leave XRBs relatively unchanged but radically alter the structure of AGN disks.  The main argument against radiation pressure driven instabilities reshaping disks is that there is no pervasive evidence  of the variability that might be expected \citep[see e.g,][]{2007A&ARv..15....1D}, but it is conceivable that such flows find an alternative equilibrium in which such variability is quenched.

A second possibility is that the much larger opacities present at the lower temperatures in AGNs due to dust and ions could significantly alter the disk structure or drive outflows.  There is considerable evidence of powerful outflows in AGN systems. Broad absorption-line quasars show blueshifted absorption lines in the optical band with velocities $\ge$ 10,000 km/s \citep{1991ApJ...373...23W} and some sources show ultrafast outflows in their X-ray spectra \citep{2010A&A...521A..57T}.  The former are thought to possibly be driven by radiation pressure on optical and UV lines \citep{1995ApJ...451..498M} whereas the latter may also be driven by radiation pressure on electrons. Numerical simulations of radiation pressure-driven outflows \citep{2004ApJ...616..688P} suggest substantial mass loss could be occurring at hundreds of gravitational radii or less, where the accretion disk continuum spectrum is formed. Such high levels of mass loss could have important implications for the accretion disk spectrum \citep{2014MNRAS.438.3024L}.

\begin{figure}[h]
\includegraphics[width=0.8\textwidth]{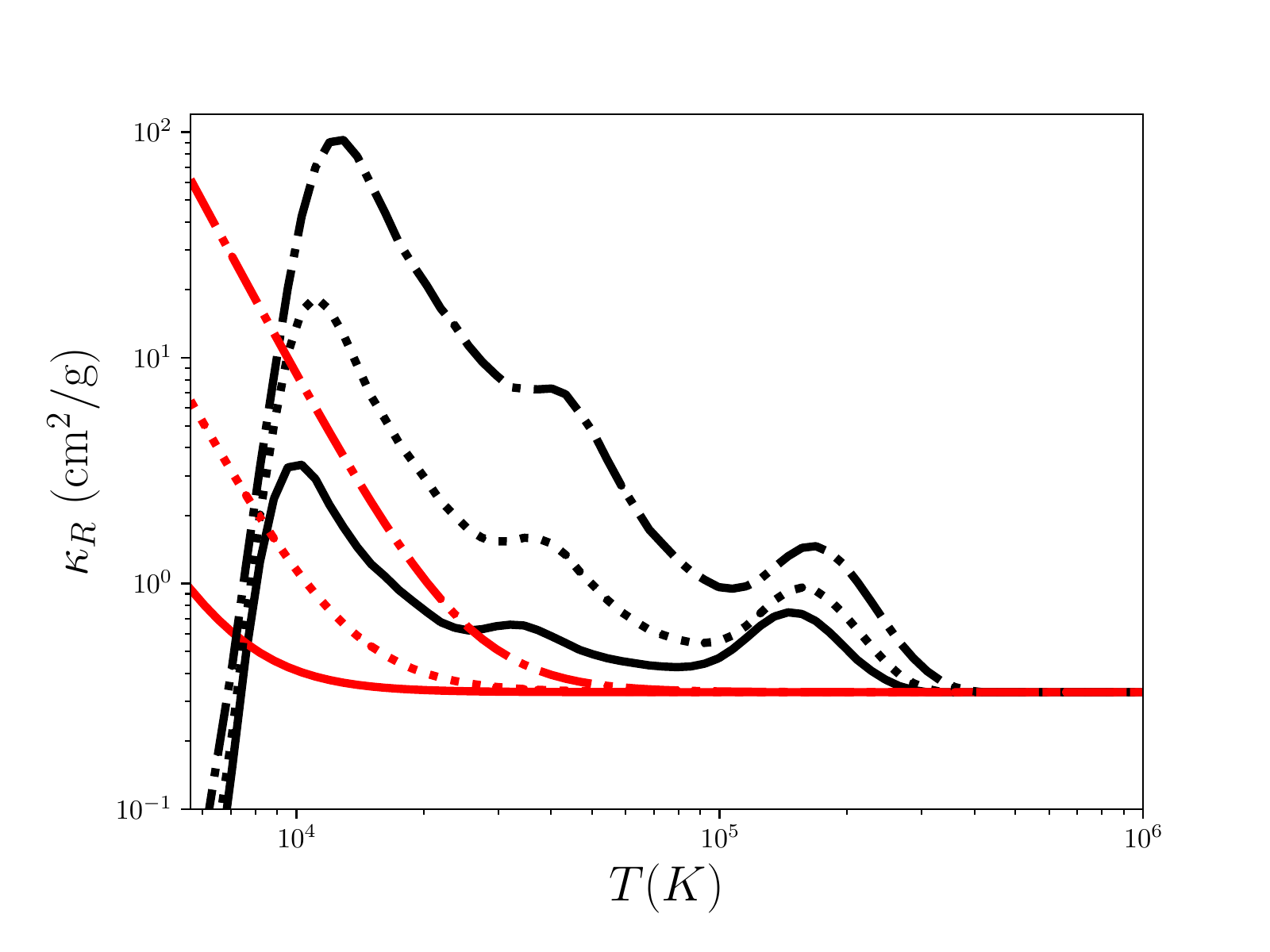}
\caption{Comparison of a sum of electron scattering opacity with Rosseland mean opacities from the OPAL project (black curves) and a Kramer's-like estimate free-free opacity (red curves). Line shapes correspond to densities of $10^{-10}$ (solid), $10^{-9}$ (dotted), and $10^{-8}$ (dash-dotted) $\rm g/cm^3$. The modest bump at $\sim 3 \times 10^5$ K is the Fe opacity bump.}
\label{f:opacity}
\end{figure}

The simulation of line-driven winds is challenging because of the need to account for the impact of X-rays on the ionization state of the gas \citep{2004ApJ...616..688P}. Even if the X-rays are modeled through radiation transfer, one still needs to use an approximate force multiplier scheme to account for the large number of bound-bound transitions that may play a role in the acceleration \citep{1975ApJ...195..157C}.  In principle, a simpler approach uses mean opacities (e.g. Rosseland or Planck mean opacities) to account for the atomic species.  The Rosseland mean is usually used to specify the radiation force, because it is the relevant opacity for a radiation-weighted mean in the diffusion limit \citep{1986rpa..book.....R}. Opacities in approximately the relevant range of AGN disks have already been constructed for computing the evolution of massive stars, and Figure~\ref{f:opacity} shows the Rosseland mean opacities from the OPAL project \citep{1996ApJ...464..943I}.

Though it looks like only a modest enhancement, the feature near $T \sim 3 \times 10^5$ K known as the Fe opacity bump is thought to produce extreme variability and help drive outflows in the envelopes of massive stars \citep{2018Natur.561..498J}. Looking at equations~(\ref{eq:teff}) and (\ref{eq:tmid}), we see that inner regions of near-Eddington accretion rate flows tend to be hotter than this in the inner-most radii of the disk, so electron scattering will dominate.  However, as one moves out, there is always a broad range of radii that are in the relevant temperature and density regime.  For black hole masses of $\gtrsim 10^6 M_\odot$ this happens in the inner several hundred gravitational radii, where the bulk of optical to UV radiation is produced.  It is conceivable that dynamics driving extreme convection and outflows in massive stars are also generically occurring in massive black holes.

A third set of considerations is the impact of the different environments and feeding of the two systems.  AGNs are thought to be fed predominantly from gas in the large-scale interstellar medium (ISM) of their host galaxies, whereas XRBs are fed by Roche lobe overflow or wind from a companion star.  Outer regions of AGN disks are expected to be self-gravitating, and either system could be subject to H ionization instabilities \citep{2001NewAR..45..449L}.  AGN accretion flows are also expected to be coincident with nuclear star clusters of their host galaxies and may have non-trivial interaction with stars or stellar remnants \citep[see, e.g.,][]{2005ApJ...619...30M,2012MNRAS.425..460M}.

Can these differences in the outer regions of the disks lead to differences in the flow near the inner radii, where most of the emission is thought to originate?  Because most of the key timescales (inflow, thermal, dynamical) decrease with radius, one might expect the disk to lose information about its large-scale feeding and slowly adjust to a similar steady state model if other considerations (e.g. opacities, radiation pressure) do not drive differences in the innermost regions.  However, this argument may not hold if the large-scale geometry of the magnetic fields is influenced by conditions in the outer flow. Evidence from numerical simulations suggests that the presence of a poloidal \begin{marginnote}[-2pt]\entry{Poloidal}{the poloidal component of magnetic field is the sum of $R$ and $z$ components}\end{marginnote}field may impact both the accretion disk and jet formation, but the presence of poloidal field may be partially dependent on whether such disks can efficiently accrete a large-scale poloidal field to their inner regions \citep{1994MNRAS.267..235L,2009ApJ...707..428B,2018ApJ...857...34Z} or generate it in situ \citep{2008ApJ...678.1180B,2012MNRAS.423.3083M,2020MNRAS.494.3656L}.

We emphasize that radiation plays a key role, particularly in the first two of these considerations and may in the third as well.  If one wants to understand the differences between AGNs and XRBs, then models and simulations in which radiation is treated explicitly are required.  The physics involved also requires dynamical and multidimensional treatments. Such considerations strongly motivate general relativistic radiation MHD numerical simulation to study the evolution of such flows with realistic opacities, initial conditions, and boundary conditions appropriate for AGNs.  Of these questions, the most difficult problem for simulations to address is the effects of differences in large-scale feeding.  The discrepancies in timescales between the largest and smallest radii are a significant impediment to numerical simulations, which must resolve the shortest flow times in the inner regions but must be run for many inflow times to achieve steady states at large radii. Nevertheless, simulations can examine what impact, if any, the magnetic field strength and geometry have on the accretion flow because the presence and strength of the poloidal magnetic field can be controlled through the initial conditions.  Hence, we think that the key questions driving numerical simulation of luminous accretion flows in AGNs are the following:

\begin{itemize}

\item What role does radiation and radiation pressure play in AGNs?  Are radiation pressure-driven instabilities present, and how do they manifest in the flow structure and dynamics?

\item Are a significant fraction of observed AGNs accreting near enough to (or above) the Eddington limit so that effects of advection and outflows become important?  How do these effects modify the light curves and SEDs of accretion flows in this limit?
  
\item What role do dust and atomic opacities play in the  dynamics and structure of AGNs?  Are there significant outflows of gravitationally bound or unbound material launched from the inner regions of AGNs even when the accretion rates are below the Eddington limit for electron scattering opacity?

\item  What role do magnetic fields play in AGNs?  Does magnetic pressure support change the structure of the disks?  Is there any reason to believe that magnetic field geometries differ between AGNs and XRBs?

\end{itemize}

Many of these effects are equally important for understanding accretion in XRBs and possibly ultraluminous X-ray sources, which may be the result of accretion above the Eddington limit.  However, here we focus on the issues specific to AGNs.

\subsection{Status of the Standard Jet Model Interpretation of AGNs}

It is generally agreed that for a black hole to launch jets it needs to be accreting. However, the ingredients necessary for jet formation are poorly understood. Is it the black hole spin or the rotation of the accretion disk that powers the jets? Complicating the situation, the ability of black hole systems to produce jets depends on the state of the accretion flow: The same black hole can exist in jetted and jet-phobic states depending on how it is fed. Perhaps the best evidence for this comes from microquasars: XRBs show radio emission that typically varies during spectral state transitions \citep{2004MNRAS.355.1105F}. Before the transition, the system is in a low/hard accretion state characterized by a nonthermal spectrum and accompanied by weak continuous jets. As the transition starts, the luminosity of the system goes up, and powerful transient jets emerge. As the luminosity declines, the jets disappear, revealing a near-thermal accretion flow. The power of transient jets appears to correlate with the black hole spin (\citealt{2012MNRAS.419L..69N,2013ApJ...762..104S}; but see \citealt{2013MNRAS.431..405R}), suggesting that black hole rotation plays a role in powering these outflows. However, how this correlation emerges is unclear.

\begin{figure}[t]
\begin{center}
\includegraphics[width=0.8\textwidth]{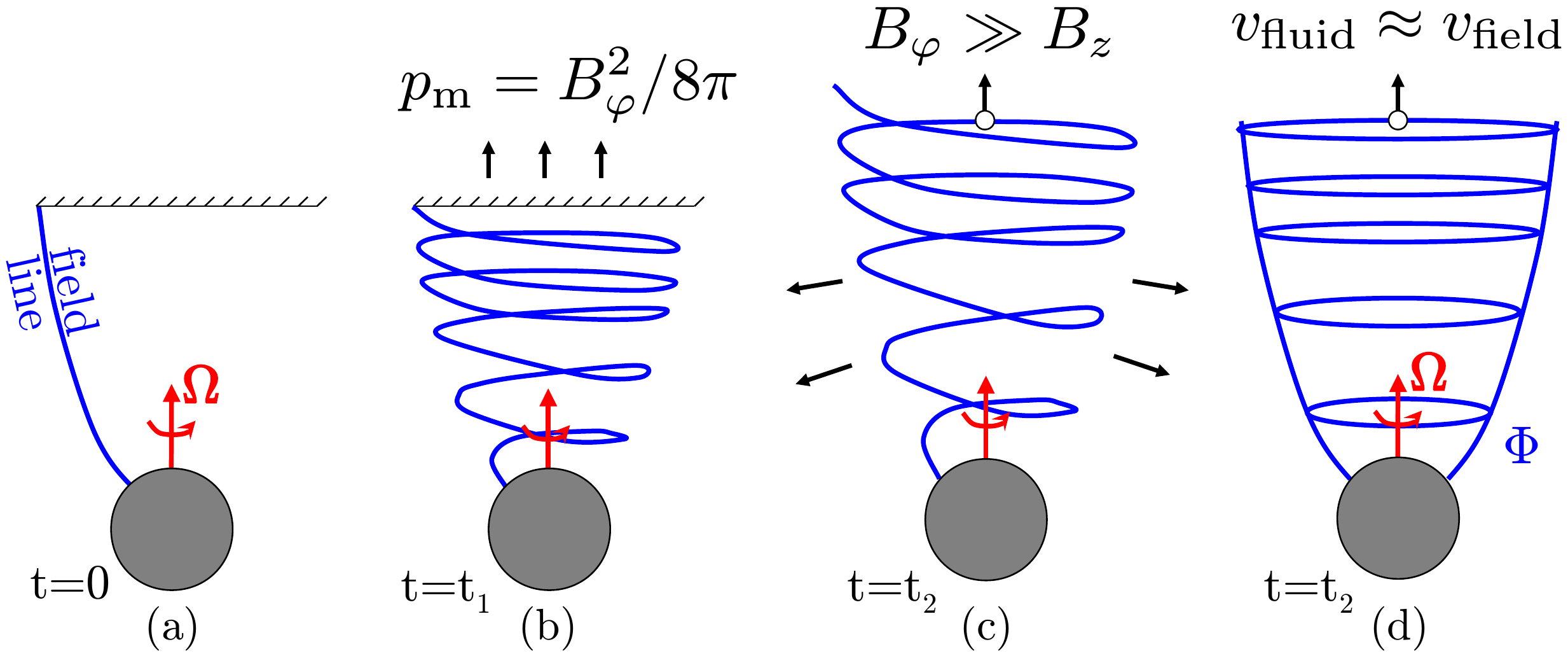}
\end{center}
\caption{\label{f:jetcartoon} Illustration of jet formation by magnetic fields.
(a)~Consider a purely poloidal (i.e., toroidal field
vanishes, $B_\varphi=0$)
field line attached on one end to a stationary ``ceiling'' (which
represents the ambient medium and is shown by a hashed horizontal
line) and on the other end to a
perfectly conducting sphere (which represents the
central black hole or neutron star and is shown by a gray filled circle)
rotating
at an angular frequency~$\Omega$. (b)~After $N$
rotations, at time $t=t_1$, the initially purely poloidal field line
develops $N$ toroidal loops.  This magnetic spring pushes against
the ceiling with an effective pressure $p_m \sim B_\varphi^2/8\pi$ due
to the toroidal field, $B_\varphi$.  As time goes on, more
toroidal loops form, and the toroidal field becomes stronger. 
(c)~At some later time, $t=t_2$, the pressure becomes so large that the
magnetic spring, which was twisted by the rotation of the sphere, 
pushes away the ceiling
and accelerates the plasma attached to it along the rotation axis,
forming a jet.  
Asymptotically far from the center, the toroidal field is the
dominant field component and determines the dynamics of the jet.
(d)~It is convenient to think of the jet as a collection of
toroidal field loops that slide down the poloidal field lines and
accelerate along the
jet under the action of their own pressure gradient and tension (hoop stress). The
rotation of the sphere continuously twists the poloidal field into
new toroidal loops at a rate that, in steady state, balances the
rate at which the loops move downstream. The power
of the jet is determined by two parameters (equation~\ref{eq:bz}): the rotational frequency
of the central object, $\Omega$, and the radial magnetic flux threading the
object, $\Phi$. Figure adapted from \citet{2015ASSL..414...45T}.
}
\end{figure}

To understand the disk-jet connection, let us first review the basics of jet formation without attempting to connect it to the disk. For simplicity, let us consider a perfectly conducting spinning sphere, shown in Fig.~\ref{f:jetcartoon}(a), which is meant to represent the central compact object (a black hole, neutron star, white dwarf, or even normal star), and a perfectly conducting ``ceiling'', which is meant to represent the ambient medium. Suppose a magnetic field line connects the sphere to the ceiling. As the sphere rotates, it coils up the field line into a magnetic spring (Fig.~\ref{f:jetcartoon}b), which pushes the ceiling away and accelerates under the action of its own pressure (Fig.~\ref{f:jetcartoon}c). One can view this acceleration in an alternative way (Fig.~\ref{f:jetcartoon}d): the rotation of the central sphere continuously reprocesses the initially vertical field line into toroidal field loops that emanate from the sphere and slide up the jet. As they do so, they expand, their pressure drops, and the pressure gradient accelerates them away. The situation for black holes, qualitatively, is rather similar. Even though a black hole does not have a physical surface, black hole rotation leads to the dragging of the inertial frames near the black hole. This causes the magnetic field lines to rotate in a surprisingly similar fashion to the perfectly conducting sphere.

\citet{1977MNRAS.179..433B} showed, in the limit of slow rotation, that a spinning black hole immersed into a large-scale vertical magnetic field would produce jets of power
\begin{equation}
  P_{\rm BZ}= k (a/2r_g)^2 \Phi_{\rm BH}^2 c,
  \label{eq:bz}
\end{equation}
where $\Phi_{\rm BH}$ is the magnetic flux threading the black hole event horizon, $-1\le a\le1$ is a dimensionless black hole spin parameter, and $k$ is a dimensionless proportionality factor. Ignoring constant prefactors, we can obtain this expression from dimensional analysis by writing that the jet power is the product of magnetic energy density, $\propto B^2$, the cross-section of the base of the jet, $\propto r_g^2$, and the speed $v\sim c$ with which the energy flows through the jets, giving $P \propto a^2 r_g^2 B^2 c$. Here, we introduce the $a^2$ prefactor to account for the variation of jet power with black hole spin (it has to be an even power of spin because spin sign change, by symmetry, leaves the power unchanged). Switching from the field strength, $B$, to the magnetic flux, $\Phi_{\rm BH} \sim B r_g^2$, gives $P\propto (a/r_g)^2 \Phi_{\rm BH}^2c$, which has the same scaling as in eq.~(\ref{eq:bz}).

With the advent of numerical simulations, it became clear that though equation~(\ref{eq:bz}) works well for small values of $a$, it underestimates the jet power for rapidly spinning black holes \citep{2001MNRAS.326L..41K}. Comparison against numerical solutions shows that replacing $r_g$ with the event horizon radius, $r_{\rm H}=r_g[1+(1-a^2)^{1/2}]$, gives a more accurate expression,
\begin{equation}
  P_{\rm jet} = k (a/r_{\rm H})^2 \Phi_{\rm BH}^2 c \times f(a),
  \label{eq:bz2}
\end{equation}
that serves well for most practical purposes at $a\lesssim0.95$ with $f(a) = 1$ \citep{2010ApJ...711...50T}. A higher-order correction, $f(a) = 1+0.35(ar_g/r_{\rm H})^2-0.58(ar_g/r_{\rm H})^4$, allows eq.~(\ref{eq:bz2}) to maintain accuracy all the way up to $a=1$ \citep{2010ApJ...711...50T,2015ApJ...812...57P}.

Thus, there appears to be a clear relationship between the black hole spin and jet power. What, then, causes the jet power to change for any given black hole? Since in AGNs the black hole spin does not change in our lifetime, changes in jet power can only come from the variations in the black hole magnetic flux, $\Phi_{\rm BH}$. To understand the disk-jet connection, we therefore need to understand how the accretion physics determines the value of $\Phi_{\rm BH}$. If we were to remove the accretion disk, then the black hole would lose its magnetic flux: By the no hair theorem, the black hole can only have three hairs -- mass, spin, and charge \citep{1973grav.book.....M}. Thus, the accretion disk holds the magnetic flux on the black hole and prevents it from slipping away: Pressure of the magnetic flux on the black hole must be in some way balanced by the pressure of the accretion flow. Here, we can again turn to the dimensional analysis and write that the magnetic pressure on the black hole is proportional to $\Phi_{\rm BH}^2$.  This balances the ram pressure in the inflow, which should be roughly proportional to the mass accretion rate, $\dot M$. To characterize the relative strength of the two, we introduce a dimensionless black hole magnetic flux, $\phi_{\rm BH} = \Phi_{\rm BH}/(\dot M r_g^2 c)^{1/2}$. Its value is set by the interaction with the disk.  The nonlinearity and intrinsic time-variability of the disk-jet interaction makes numerical simulations an attractive approach for constraining the allowed range of values of $\phi_{\rm BH}$. As we discuss in Section~\ref{sec:simulations-agn-jets}, the allowed range of $\phi_{\rm BH}$ spans the range from zero to a maximum value around 50 for which the black hole magnetic flux becomes as strong as the gravity that keeps the disk on an orbit around the black hole \citep{2011MNRAS.418L..79T}.\begin{marginnote}[-2pt]\entry{Magnetically arrested disk (MAD)}{accretion state where magnetic flux inside the disk becomes strong enough to disrupt the flow}\end{marginnote} In this magnetically arrested disk (MAD) state, the magnetic flux is as strong as possible and weakly depends on the black hole spin, $\phi_{\rm MAD} \approx 70(1-0.38ar_g/r_{\rm H})h_{0.3}^{1/2}$, where $h = r\times 0.3h_{0.3}$ is the half-thickness of the disk. This corresponds to the following maximum jet power:
\begin{equation}
  \label{eq:pmad}
  P_{\rm MAD} = \eta_{\rm MAD} \dot M c^2 \approx 1.3 h_{0.3} a^2 \dot M c^2,
\end{equation}
where we used the fact that the jet efficiency is $\eta_{\rm MAD} = k\phi_{\rm MAD}^2 a^2 (r_g/r_{\rm H})^2 f(a) \approx 1.3 a^2h_{0.3}$  \citep{2015ASSL..414...45T}. Thus, the jet power $P_{\rm jet}$ takes on a value between $0$ and $P_{\rm MAD}$ depending on the strength of the black hole magnetic flux $\phi_{\rm BH}$ relative to its maximum possible value, $\phi_{\rm MAD}$:
\begin{equation}
  \label{eq:pjet}
  0\le P_{\rm jet} = \left(\frac{\phi_{\rm BH}}{\phi_{\rm MAD}}\right)^2 P_{\rm MAD}\le P_{\rm MAD}= 1.3 h_{0.3}a^2\dot{M}c^2.
\end{equation}

Now that we understand the jet power, there are still many questions left unanswered. These questions are driving detailed studies in various branches of astrophysics include the following:
\begin{itemize}
  \item Where does the magnetic flux powering jets come from? Does it need to be dragged inward from the ISM or can it be generated inside the disk in situ?
  \item How are transient jets launched during spectral state transitions in XRBs? Does an analog of the state transitions and transient jets exist in AGNs?
  \item How do jets accelerate to relativistic velocities? Is radiation pressure important in launching and accelerating the jets?
  \item Do jets heat up the interstellar gas and affect galaxy evolution? Can jet feedback lead to the $M-\sigma$ relationship?
  \item What makes jets shine? What can we learn from the observations of the black hole shadow with the Event Horizon Telescope? What can the jets tell us about the strong-field gravity and general relativistic frame dragging that birthed them?
\end{itemize}
We discuss several of these questions in Sections~\ref{sec:overv-accr-disk} and \ref{sec:simulations-agn-jets}.

\section{OVERVIEW OF ACCRETION DISK AND JET SIMULATION METHODS}
\label{sec:overv-accr-disk}

\subsection{General Relativistic MHD}

The  numerical simulation of accretion flows onto black holes has a long history, with \cite{1972ApJ...173..431W} already considering the flow of fluid with nonzero angular momentum in the Kerr spacetime. Early work primarily focused on hydrodynamics models \citep[e.g.,][]{1984ApJ...277..296H}, but the realization that the MRI could provide the necessary angular momentum transport led to the development and application of MHD methods \citep{1998RvMP...70....1B}. With this realization it became standard for numerical simulations to solve some version of the MHD equations, which includes conservation of mass,
\begin{eqnarray}
  \frac{\partial\rho}{\partial t}+\bfnabla\cdot(\rho \bv) = 0,\label{eq:mass}
\end{eqnarray}
conservation of momentum,
\begin{eqnarray}
  \frac{\partial( \rho\bv)}{\partial t}+\bfnabla\cdot({\rho \bv\bv-\bb\bb+{{\sf P}^{\ast}}}) = -\rho\bfnabla\Phi,
  \label{eq:momentum}
\end{eqnarray}
conservation of energy,
\begin{eqnarray}
  \frac{\partial{E}}{\partial t}+\bfnabla\cdot\left[(E+P^{\ast})\bv-\bb(\bb\cdot\bv)\right] = -\rho\bv\cdot\bfnabla\Phi, \label{eq:energy}
\end{eqnarray}
and the induction equation,
\begin{eqnarray}
\frac{\partial\bb}{\partial t}-\bfnabla\times(\bv\times\bb) = 0.
\label{eq:induction}
\end{eqnarray}
Here, $\rho$, $\bb$, $\bv$ are density, magnetic field and flow velocity, ${\sf P}^{\ast}\equiv(P_g+B^2/2){\sf I}$ (with ${\sf I}$
the unit tensor), $P_g$ is the gas pressure, and the magnetic permeability $\mu=1$.  The total gas energy density is
\begin{eqnarray}
E=E_g+\frac{1}{2}\rho v^2+\frac{B^2}{2},
\end{eqnarray}
and $\Phi$ is the gravitational potential.  The precise form of these equations varies depending on application.  Sometimes the energy equation is omitted and an isothermal equation of state is adopted.  Additional source terms may be added to the right hand sides of the equations of momentum and energy, such as the effects of radiative cooling, heating, or radiation forces.  This may require the solution of additional equations, such as the radiation transfer equation or its moments.

\begin{textbox}[ht]
\section{Numerical Methods for Solving the MHD Equations}
There are a significant number of different methods available for simulating MHD flows and an even larger number of codes that implement them.  Most methods for solving fluid dynamics equations fall into two categories: Eulerian or Lagrangian.  Eulerian methods tend to solve finite difference representations of the equations on a fixed mesh.  Finite volume (Godunov) methods \citep{lev02} are particularly popular for their shock-capturing capabilities because shocks commonly arise in astrophysical flows.  Lagrangian methods tend to utilize particle-based representations with fluid properties advected along with the particles, such as smoothed particle hydrodynamics \citep{sphbook}.  Other examples include moving mesh codes that carry a time variable mesh along with the particles or fluid flow \citep[e.g.,][]{2010MNRAS.401..791S,2011ApJS..197...15D}.

Both Eulerian and Lagrangian codes have their merits, but the majority of the simulations discussed in this review utilize mesh-based Eulerian schemes.  This is partly due to their shock-capturing capabilities and their simplicity relative to moving mesh methods.  Another major consideration is the need to preserve the divergence-free nature of the magnetic field: $\nabla \cdot \mathbf{B} = 0$.  Although it should always hold in nature, this condition is not preserved by all integration methods.  Several codes address this issue by allowing the divergence to develop but then limiting it via a divergence cleaning scheme.  Other integration schemes are specifically designed to preserve the divergence-free condition to machine precision. A popular example is the constrained transport scheme \citep{1988ApJ...332..659E}, which is most easily implemented on structured Eulerian mesh \citep[c.f.][]{2014MNRAS.442...43M}.\end{textbox}

Early MHD simulations of accretion focused on shearing box simulations \citep{1995ApJ...440..742H,1995ApJ...446..741B,1996ApJ...463..656S}, which adopt shearing boundary conditions and add the Coriolis force to mimic the effects of fluid rotation in an accretion disk but only simulate a small patch of the disk.  Although these simulations are still widely used to study the structure, dynamics and thermodynamics of the disk at high resolution, concerns about convergence with resolution \citep{2007ApJ...668L..51P,2007A&A...476.1113F,2017ApJ...840....6R} and dependence on box size \citep[e.g.][]{2012MNRAS.422.2685S,2016MNRAS.456.2273S} remain. (See the sidebar titled Numerical Methods for Solving the MHD Equations.)

Global simulations of MHD black hole accretion flows were first performed with pseudo-Newtonian potentials \citep{2002ApJ...573..738H} but these were supplanted by general relativistic MHD (GRMHD) simulations of accretion flows onto spinning and nonspinning black holes \citep{2003ApJ...589..444G,2003ApJ...599.1238D}. In GRMHD, the covariant generalizations of the MHD equations are conservation of mass, stress-energy, and the relativistic Maxwell's equations, which are evolved on a choice of coordinates (usually Boyer-Lindquist or Kerr-Schild) dictated by the Kerr spacetime. \begin{marginnote}[-2pt]\entry{GRMHD}{General relativistic magnetohydrodynamics}\end{marginnote} The first GRMHD simulations lacked realistic cooling and, therefore, best approximate low-luminosity accretion flows.  Nevertheless, they have led to a much better understanding of flow in the innermost regions near the black hole, including accretion in the plunging region, driving of outflows, and launching of jets.

In particular, it became clear that in the presence of large-scale vertical magnetic flux jets are a typical outcome of radiatively inefficient, geometrically thick black hole accretion disks with $h/r\sim 0.3{-}1$ \citep{2004ApJ...611..977M,2005ApJ...620..878D,2005ApJ...630L...5M,2006ApJ...641..103H}. Radiatively efficient, geometrically thin disks with $h/r\lesssim0.05$ were simulated via the inclusion of cooling terms that kept the disk thickness at a desired value \citep{2008ApJ...687L..25S,2010MNRAS.408..752P,2010ApJ...711..959N}. These simulations did not produce jets, in agreement with the observations that show no detectable radio emission (and, hence jet activity) from thin disks in XRBs and the vast majority (90\%) of quasars. The remaining 10\% of quasars show powerful radio jets whose origins are still a mystery. Recent GRMHD simulations show that even very thin disks, $h/r\sim0.015$, are capable of producing powerful jets and provide potential resolution to this puzzle (see Sec.~\ref{sec:disk-jet-connection} and \citealt{2019MNRAS.tmp..813L}).

Interestingly, it appears that higher resolutions than typical (e.g., $256^3$ cells) are required to achieve convergence in the value of the effective $\alpha$-viscosity parameter in global GRMHD simulations of weakly magnetized accretion disks \citep{2019ApJS..243...26P}. Global simulations of strongly magnetized accretion disks are less sensitive to resolution \citep{2019ApJ...874..168W}.

\subsection{Radiation Hydrodynamics}
Adding and improving radiation transfer treatments in numerical simulations of accretion flows is an important focus of recent work.  The relevant equation to solve is the radiative transfer equation,
\begin{equation}
\frac{1}{c} \frac{\partial I_\nu}{\partial t}+\mathbf{n}\cdot\mathbf{\nabla } I_\nu= \eta_\nu -\alpha_\nu I_\nu,\label{eq:transfer}
\end{equation}
where $I_\nu$ is the specific intensity, $\eta_\nu$ is the emissivity, $\alpha_\nu = \kappa_\nu \rho$ is the extinction coefficient, and $\kappa_\nu$ is the opacity.  The form on the right-hand side is general if extinction includes scattering opacity contributions and the emissivity accounts for scattered radiation. Because $I_\nu$ is a function of position, angle, and frequency, it is usually computationally expensive compared to the standard MHD equations, which only depend on position. For this reason, some treatments focus on its angle and frequency-integrated moments corresponding to conservation of radiation energy,
\begin{equation}
  \frac{\partial E_r}{\partial t}+\mathbf{\nabla} \cdot \mathbf{F_r} =  c \rho
  \left (\kappa_P a T^4 - \kappa_E Er\right),\label{eq:radenergy}
\end{equation}
and conservation of radiation momentum,
\begin{equation}
  \frac{1}{c^2} \frac{\partial \mathbf{F_r}}{\partial t}+\mathbf{\nabla} \cdot {\sf P_r} =
  -\frac{\kappa_F \rho}{c} \mathbf{F_r}.\label{eq:radmomentum}
\end{equation}
Here, $E_r$, $\mathbf{F_r}$, and $\sf P_r$ are the radiation energy density, flux, and pressure tensor, respectively.  The opacities $\kappa_P$, $\kappa_E$, and $\kappa_F$ correspond to the Planck, energy, and flux mean opacities.  In most applications, the Rosseland mean is used for $\kappa_F$ and the Planck mean is used for $\kappa_E$.

For the sake of brevity, we will summarize some of the most salient points and refer the reader to \citet{1984oup..book.....M} for a comprehensive discussion of theses equations.  Although these equations look simple, one should note that, as written, all variables are in the Eulerian frame, but the simple isotropic forms for the emissivities and opacities only hold in the comoving frame. Hence, one usually must Lorentz transform the source terms, expand the right-hand sides in powers of $v/c$, or use a comoving frame approach which introduces additional acceleration terms.  Compton scattering also generally introduces additional terms.  The source terms on the right hand side of equations~(\ref{eq:radenergy}) and (\ref{eq:radmomentum}) represent the transfer of energy and momentum from the radiation field to fluid.  Hence, the negative of these terms are added to the right-hand side of equations~(\ref{eq:momentum}) and (\ref{eq:energy}), corresponding to net heating/cooling by the radiation field and the radiation force.

Early efforts to include radiation predominantly focused on simulations utilizing flux-limited diffusion \citep[FLD;][]{1981ApJ...248..321L}.\begin{marginnote}[-2pt]\entry{Flux-limited diffusion (FLD)}{diffusion-like approximation to radiation transfer}\end{marginnote} In FLD, only equation~(\ref{eq:radenergy}) is retained, and $\mathbf{F_r}$ is computed from the gradient of $E_r$ using a diffusion approximation. A limiter is utilized to keep $|\mathbf{F_r}| \le c E_r$ in the optically thin regime. This method has been used in the context of shearing box \citep{2004ApJ...605L..45T} and global accretion flow \citep{2005ApJ...628..368O} simulations. More recently, several groups have developed general relativistic radiation transfer modules based on two moment methods with M1 closure \citep{2013MNRAS.429.3533S,2014MNRAS.441.3177M,2016MNRAS.463.3437M}.\begin{marginnote}\entry{M1}{approximate closure scheme used relate radiation pressure tensor to the radiation flux and energy density}\end{marginnote}  In M1 closure schemes, equations~(\ref{eq:radenergy}) and (\ref{eq:radmomentum}) or their relativistic generalization (corresponding to conservation of radiation stress-energy) are solved.  Because $\sf P_r$ appears in equation~(\ref{eq:radmomentum}), it must be specified. This is generally done by introducing a new tensor called the Eddington tensor ${\sf f}\equiv {\sf P_r}/E_r$ which characterizes the angular distribution of the radiation field.  In the M1 scheme, $\sf f$ is computed as a function of $E_r$ and $\mathbf{F_r}$.  

Other approaches include direct solution of the angle-dependent transfer equation \citep{2014ApJ...784..169J} or the variable Eddington tensor (VET) method \citep{2012ApJS..199...14J}.\begin{marginnote}\entry{variable Eddington tensor (VET)}{closure scheme used compute radiation pressure tensor from an approximate solution of the transfer equation.}\end{marginnote}  In the direct solution approach, a frequency-averaged version of equation~(\ref{eq:transfer}) is solved explicitly for a fixed number of angles in each zone.  In the VET method, equations~(\ref{eq:radenergy}) and (\ref{eq:radmomentum}) are integrated directly, but a time-independent solution of the transfer equations is used to compute the Eddington tensor. Becasue both the FLD and M1 methods make assumptions about the angular distribution of the radiation field by prescribing relations among $E_r$, $\mathbf{F_r}$ or $\sf P_r$, methods involving direct solution of the transfer equation are generally thought to be more accurate. The trade-off is that they are also more computationally expensive and are algorithmically more complex to implement. For this reason, incorporating general relativity in the radiation transfer equation is significantly more challenging, so simulations performed with these methods have all been nonrelativistic MHD-based, whereas many of the simulations performed with M1 methods have been general relativistic MHD-based. The formalism for the general relativistic transfer equation is well-developed \citep{2013PhRvD..88b3011C} and efforts to implement it in black hole accretion simulations are ongoing.

\section{MHD SIMULATIONS OF AGN ACCRETION DISKS}

To date, there have been relatively few simulations directly aimed at studying supermassive black hole accretion disks.  Prior to the addition of radiation transfer in simulations, there was usually no explicit dependence on the black hole mass in the simulations.  This is because the bare MHD equations can be arbitrarily rescaled by an explicit choice of length scale or characteristic density. Choosing the characteristic length scale specified the mass, and choosing the density then specified the mass accretion rate. Hence, one could scale the simulation to the supermassive black hole regime, but there was nothing to distinguish the dynamics or thermodynamics of the rescaled simulations from that of stellar mass black holes.  However, this rescaling freedom is lost when radiation is added, because the dependence of opacities and emissivities on temperature and density enforces an explicit choice of length and density scales.  In this sense, radiation transfer is fundamental to the distinction between supermassive and stellar mass black holes.  Although we have learned a great deal about accretion flows from pure MHD simulations, they cannot tell us about the differences between AGN flows and XRB flows, so we choose to focus on radiative simulations in this section of the review.

The first sets of radiation MHD simulations of accretion flows utilized the FLD approximation \citep{1981ApJ...248..321L,2001ApJS..135...95T} to perform local shearing box simulations \citep[e.g.][]{2003ApJ...593..992T,2004ApJ...605L..45T,2006ApJ...640..901H}.  With the exception of \citet{2004ApJ...605L..45T}, these simulations focused on patches of disk around $\sim 10 M_\odot$ black holes, where the radiation-to-gas pressure ratio tended to be lower.  At the same time, efforts were underway to use FLD to study the global structure of accretion flows.  Initially, this took the form of two-dimensional, viscous hydrodynamic flows \citep{2005ApJ...628..368O}, but was eventually generalized to MHD treatments \citep{2011ApJ...736....2O}.  These calculations were mainly aimed at studying the variation of the flow with accretion rate and also focused on $10 M_\odot$ black holes.

\subsection{Thermal Instability in Radiation Pressure Dominated Accretion}

These radiative MHD shearing box simulations were particularly interesting for testing the predictions of thermal instability in radiation pressure-dominated accretion flows \citep{1976MNRAS.175..613S}.  The initial results from the Zeus FLD simulations confirmed that gas pressure-dominated disks were stable \citep{2006ApJ...640..901H}, as expected. More intriguingly, simulations found evidence for thermal stability even when radiation pressure dominates gas pressure \citep{2004ApJ...605L..45T,2009ApJ...691...16H}.  However, these results were contradicted by Athena simulations utilizing the VET method \citep{2013ApJ...778...65J}, which generally found runaway heating or collapse.  Simulations using the Athena FLD module also showed indications of thermal runaway, so the discrepancy was not purely a result of the radiation transfer algorithm.  Instead, \cite{2013ApJ...778...65J} attribute the discrepancy to the radial width of the Zeus simulation box being too small, finding that the simulations became more stable as they narrowed the radial dimensions of the simulation domain.

Despite reaching different conclusions about stability, these simulations all found that the disks could survive for many thermal timescales, which is indicative of a possible robustness against the standard thermal instability.  Although the relation of the time-averaged stress in MHD simulations is roughly consistent with the $\alpha$-disk assumption, the shearing boxes differed from the standard model assumptions in important ways.  Magnetic pressure is not included in standard disk models, but in simulations it plays an important role in supporting the surface regions against gravity.  Thermodynamics is more complicated in MHD runs, with a greater fraction of dissipation occurring near or above the photosphere, leading the density to drop faster with height than in standard disk models. And the time dependence of the relation between central pressure and stress is less direct, so that increases (decreases) in radiation pressure did not immediately result in increases (decreases) in the scale height or dissipation \citep{2009ApJ...691...16H}.  

\cite{2004ApJ...605L..45T} was the sole simulation among those above that was run for supermassive black hole conditions, but it only included free-free (bremsstrahlung) and electron-scattering opacity.  \citet{2016ApJ...827...10J} explored the role of opacity in such environments, using the Athena VET module to simulate several shearing boxes for conditions around a $5\times10^8 M_\odot$ black hole while including OPAL opacities.  Despite much higher ratios of radiation to gas pressure compared with the lower-mass black holes discussed by \cite{2013ApJ...778...65J}, the resulting simulations appeared to be thermally stable.  In contrast, simulations performed with the same parameters, but only including free-free and electron-scattering opacity, all showed runaway collapse on very short timescales, much faster than in the lower black hole mass runs.  \citet{2016ApJ...827...10J} attribute the enhancement in stability to a combination of two effects. First, UV opacities introduce an anticorrelation in the optical depth to the midplane and central temperature, opposite to the electron scattering-dominated assumption.  Second, the opacity drives convection that isn't present in the previous simulations, which enhances the nonradiative fluxes carried in the simulations.  However, the dependence of shearing box simulations on box size, combined with time-dependent global modeling \citep{2017ApJ...845...20G} strongly motivates further study of UV opacity effects in global numerical simulations.

\subsection{Global Simulations of AGN Accretions Flows}

Local (shearing box) simulations are useful for studying accretion physics at high resolution, but one ultimately requires global simulations to obtain a more complete picture of the accretion flow.  Furthermore, the apparent dependence of thermal instability on radial width (or aspect ratio) in shearing box simulations motivates global simulations for which this artificial constraint is removed.  As already noted, global simulations of black hole accretion have advanced significantly over the past two decades, advancing from pseudo-Newtonian MHD \citep{1995ApJ...440..742H} to full GRMHD \citep{2003ApJ...589..444G}, sometimes employing ad hoc cooling function to maintain a thin disk \cite[e.g.,][]{2010MNRAS.408..752P,2010ApJ...711..959N}.  Without radiation transfer and realistic opacities, such simulations can be scaled to arbitrary masses.  Therefore, these simulations apply to both the solar mass and the supermassive black hole regime.  However, in practice, such simulations have been primarily compared with XRBs \citep[e.g.][]{2011MNRAS.414.1183K,2013ApJ...769..156S}.

For these reasons, AGN specific MHD numerical simulations are relatively recent and relatively few in number.  Nevertheless, there are several lines of related inquiry that are worth summarizing. Although their simulation was calibrated for solar mass black holes, \cite{2016MNRAS.463.3437M} performed the first radiation GRMHD simulation of radiation pressure-dominated global disk. Consistent with the earlier shearing box simulations results, they found a runaway collapse of the disk, approximately on the thermal timescale.  In contrast, \citet{2016MNRAS.459.4397S} performed global radiation GRMHD simulations in a similar regime that was stable to thermal and inflow instabilities.

Thus, we have two sub-Eddington accretion rate simulations that provide different predictions about the stability of accretion flows in this radiation pressure dominated regime.  \citet{2016MNRAS.459.4397S} attribute the stability of their simulation to magnetic pressure from toroidal magnetic fields, which is broadly consistent with predictions of earlier analytic models that suggested magnetic support could impact stability \citep{2007MNRAS.375.1070B,2017MNRAS.464.2311B}. The difference between the \citet{2016MNRAS.459.4397S} and 
\citet{2016MNRAS.463.3437M} results can then be attributed to the magnetic field geometry. The required toroidal fields are produced by threading the disk with a large poloidal flux that then gets wound up by the shear in the accretion flow; this is a picture that is broadly supported by other recent simulation
results \citep{2020MNRAS.492.1855M}, including some in the AGN regime discussed below \citep{2019ApJ...885..144J}.  These results suggest that magnetic support can stabilize accretion flows, as long as sufficiently large poloidal magnetic fields can be accreted into the inner regions of the accretion disk.

Radiation hydrodynamic simulations of line-driven AGN outflows \citep[e.g.,][]{2000ApJ...543..686P} have also been used to model properties of broad absorption line quasars.  Because these calculations treat the accretion disk and its emission as a fixed boundary condition, they cannot evolve the disk self-consistently, but it is notable that some simulations have mass outflow rates that are a substantial fraction of the assumed inflow rate.  Taken together, these simulations suggest that the impact of atomic opacities and the enhanced radiation pressure in AGNs may lead to distinct differences from XRB accretion flows.

Radiation GRMHD simulations of the radiatively inefficient accretion flow regime have also advanced significantly in the past several years, motivated by the imaging of Sgr A* and M87 with the Event Horizon Telescope \citep{2018ApJ...864..126R,2019ApJ...875L...1E}.  Because such flows are relatively optically thin, they are well suited to Monte Carlo-based algorithms for radiation transfer, such as the BHLIGHT code \citep{2015ApJ...807...31R}.  However, the standard implementation of these algorithms has not traditionally scaled well to the optically thick regime of luminous accretion flows, so these techniques have not been applied to luminous AGNs.

To date, only a few radiation MHD simulations have been performed with AGN masses and UV appropriate opacities.  These AGN-specific simulations have been performed for super-Eddington \citep{2019ApJ...880...67J} and sub-Eddington \citep{2019ApJ...885..144J} regimes using the Athena++ radiation MHD code with a pseudo-Newtonian potential. The simulations are initialized with a torus centered at 50-80 gravitational radii, seeded with a weak magnetic field.  The torus is unstable to the MRI, generating turbulence that allows an accretion flow to self-consistently form at radii interior to the torus.

The generic properties of the accretion flow are broadly similar to the two-dimensional viscous radiation hydrodynamic simulations \citep{2005ApJ...628..368O,2018PASJ...70..108K} as well as radiation GRMHD simulations performed for solar mass black holes \citep{2015MNRAS.454L...6M,2015MNRAS.447...49S,2016MNRAS.456.3929S} or super-Eddington accretion in tidal disruption events \citep[TDEs;][]{2018ApJ...859L..20D}.  At the highest accretion rates, the accretion disk is geometrically thick, with strong radiation pressure-dominated outflows at velocities of 10-30\% of the speed of light.  However, in contrast to previous simulations and slim accretion disk models, the radiative efficiencies can remain relatively high, as much as $\sim 5$\%. The highest efficiencies are consistent with earlier Athena simulations in cylindrical geometry \citep{2014ApJ...796..106J} around solar mass black holes, which attributed the enhanced radiative efficiency to vertical advection of energy by the MHD turbulence.  At higher accretion rates $(\sim 500 L_{\rm Edd}/c^2)$, radiative efficiencies drop to $\sim 1$\%, in better agreement with slim disk models and the radiation GRMHD simulations of \citet{2016MNRAS.456.3929S}.  Kinetic efficiencies are generally lower than in those GRMHD simulations, but this may be attributed (at least in part) to the use of the pseudo-Newtonian potential and the lack of a BZ process driven by a spinning black hole.

Perhaps the most intriguing results from the \citet{2019ApJ...880...67J} simulations is the excitation of spiral density waves and their contribution to driving accretion via Reynolds stresses, which in some (but not necessarily all) cases may be the dominant contribution.  This enhanced role of density waves is attributed to radiative damping of MRI turbulence in the simulations, although the details of the excitation mechanism are not yet well understood.  Radiation viscosity \citep{1971PASJ...23..425M,1984oup..book.....M} is also more important than previous thought.  Radiation viscosity acts on shear flows in the same way as standard viscosity but with photons providing the momentum exchange across the shear layer. In the super-Eddington simulations, it plays a surprisingly significant but ultimately subdominant role in the overall transport of angular momentum. The large radiation energy density enhances the role of radiation viscosity because more photons are present to transfer momentum.  However, the large optical depths and correspondingly short mean free paths between electron scattering means there are only small changes in fluid velocity differences between scatterings.

The sub-Eddington simulations \citep{2019ApJ...885..144J} offer further surprises.  Despite having accretion rates as low as 7\% of the Eddington accretion rate, these simulations appear to reach a steady state out to $\sim 20$ gravitational radii and last for a duration of many thermal timescales without any evidence for thermal or inflow instabilities.  As shown by \citet{2016MNRAS.459.4397S}, the stability is attributed to magnetic pressure support.  Although magnetic pressure is lower than radiation pressure, it has a smaller scale height and provides the dominant support of the gas against tidal gravity.  The radiation pressure scale height is large because radiation pressure within the disk is roughly constant, due to the dissipation primarily occurring in surface layers with relatively low optical depth.  Such magnetic pressure support has been proposed to aid in explaining the size and timescale discrepancies discussed in Section~\ref{theorystatus} \citep{2019MNRAS.483L..17D}.

\begin{figure}
    \begin{subfigure}{.5\textwidth}
    \centering
    \includegraphics[width=2.5in]{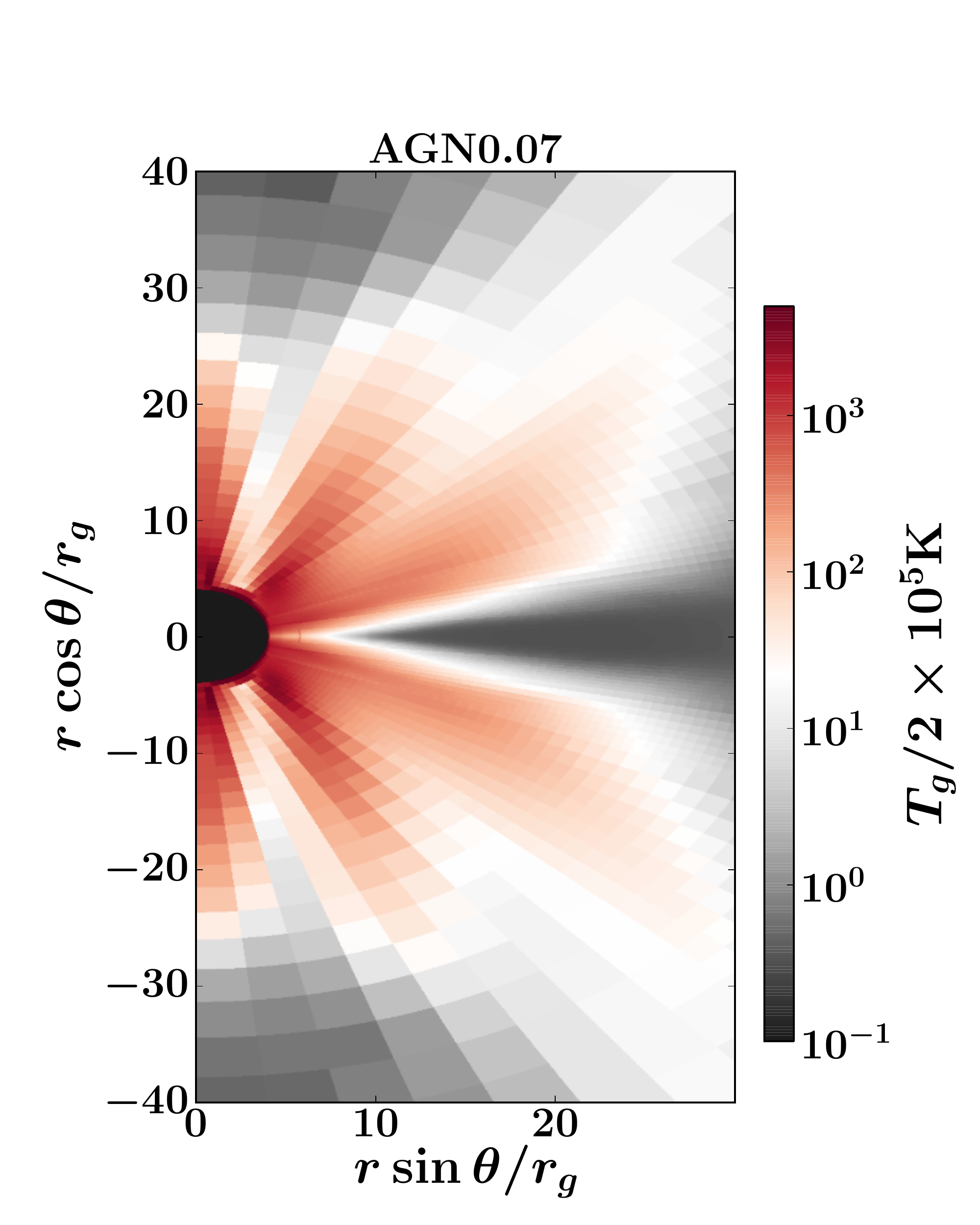}
  \end{subfigure}
  \begin{subfigure}{.5\textwidth}
    \centering
    \includegraphics[width=2.5in]{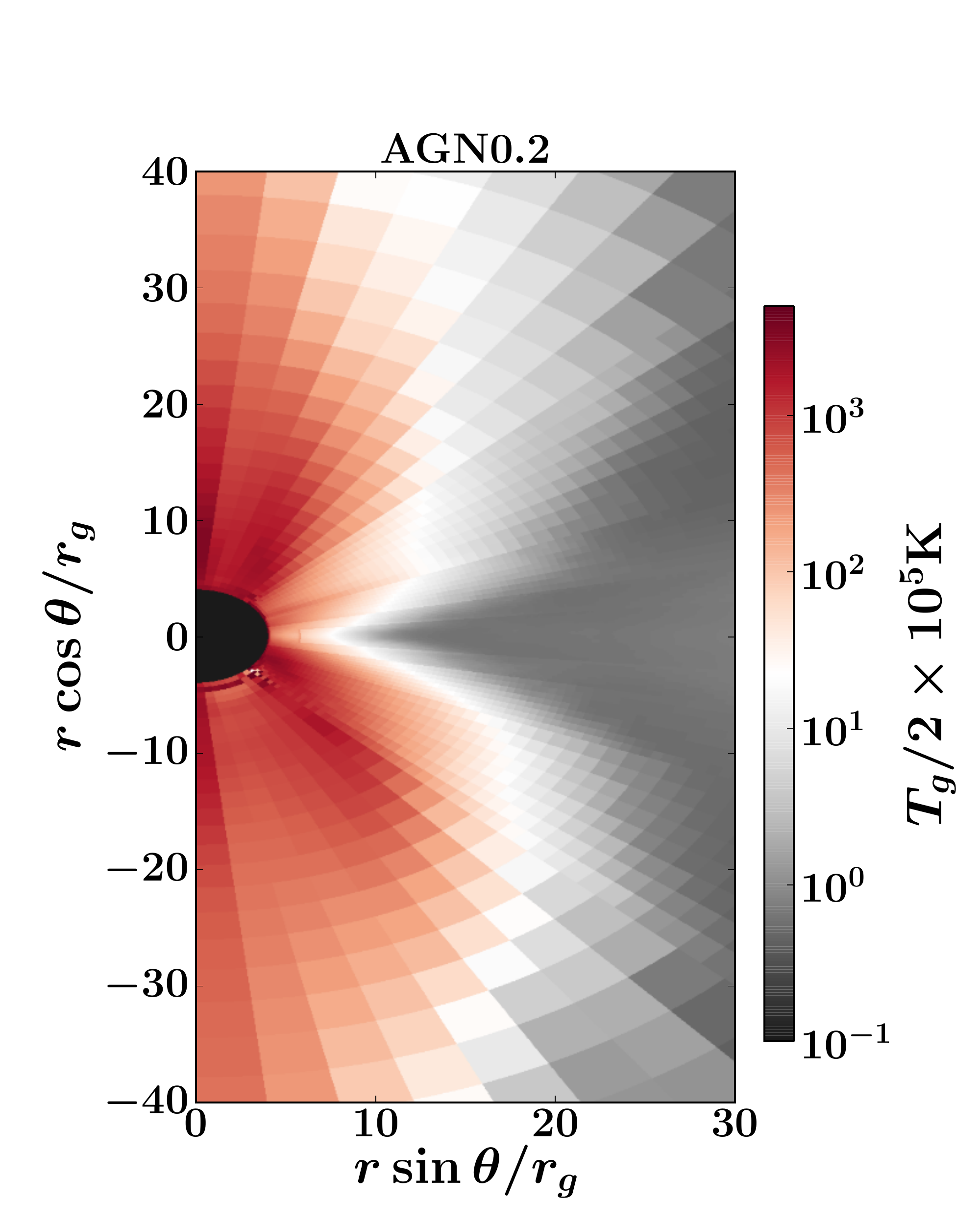}
  \end{subfigure}
\caption{Time and azimuthally averaged spatial distributions of gas temperature $T_g$ for the two simulations runs onto $5\times 10^8$ black holes from \citet{2019ApJ...885..144J}. The left and right panels have accretion rates of 7\% ({\sf AGN0.07)} and 20\% ({\sf AGN0.2}), respectively. The gas temperature is $\approx 10^5-2\times 10^5$ K in the optically thick part of the disk but rapidly increases to $10^8-10^9$ K in the optically thin coronal regions. Note how the corona is more radially extended in the lower accretion rate simulation than in the higher accretion-rate case.}\label{f:corona}
\end{figure}

The large dissipation in surface layers is the result of strong radiative viscosity. In the simulation with 7\% of the Eddington accretion rate, the radiation viscosity is actually the dominant mechanism for angular momentum transport, and more mass flows inward in the surface layers than in the midplane of the disk where MRI turbulence dominates.  The large radiation viscosity is the result of a combination of large photon mean free paths in these surface regions and large radiation energy density.

As shown in Figure~\ref{f:corona} the large surface dissipation leads to strong temperature inversions above the disk, reminiscent of models of AGN coronae.  The simulations have both temperatures high-enough to potentially produce the hard X-ray coronae and more moderate temperatures needed to explain the presence of soft X-ray excesses observed in many AGNs \citep{1993A&A...274..105W,2018MNRAS.480.1247K}.  Intriguingly, the fraction of emission in optically thin regions is higher in the lower accretion rate simulations, leading to higher temperatures consistent with the observed relation that optical to X-ray spectral indexes get harder as luminosity decreases \citep{2006AJ....131.2826S}.  Although these results are promising, we must keep in mind that all these effects are occurring in close vicinity to the black hole, where relativistic effects are important, so we must ultimately test these conclusions with general relativistic calculations.

Although OPAL opacities are used in the Athena++ simulations of super-Eddington and sub-Eddington disks, they have relatively little impact on the resulting dynamics.  This is because the inner regions of the accretion rate, where the flow reaches a steady state, are too hot for there to be significant enhancements in the opacity above electron scattering.  Future simulations need to focus on larger radii to see the effects of such opacities.

\section{SIMULATIONS OF AGN JETS}
\label{sec:simulations-agn-jets}

\subsection{The Disk-jet Connection}
\label{sec:disk-jet-connection}

\begin{figure}[t]
\includegraphics[width=\textwidth]{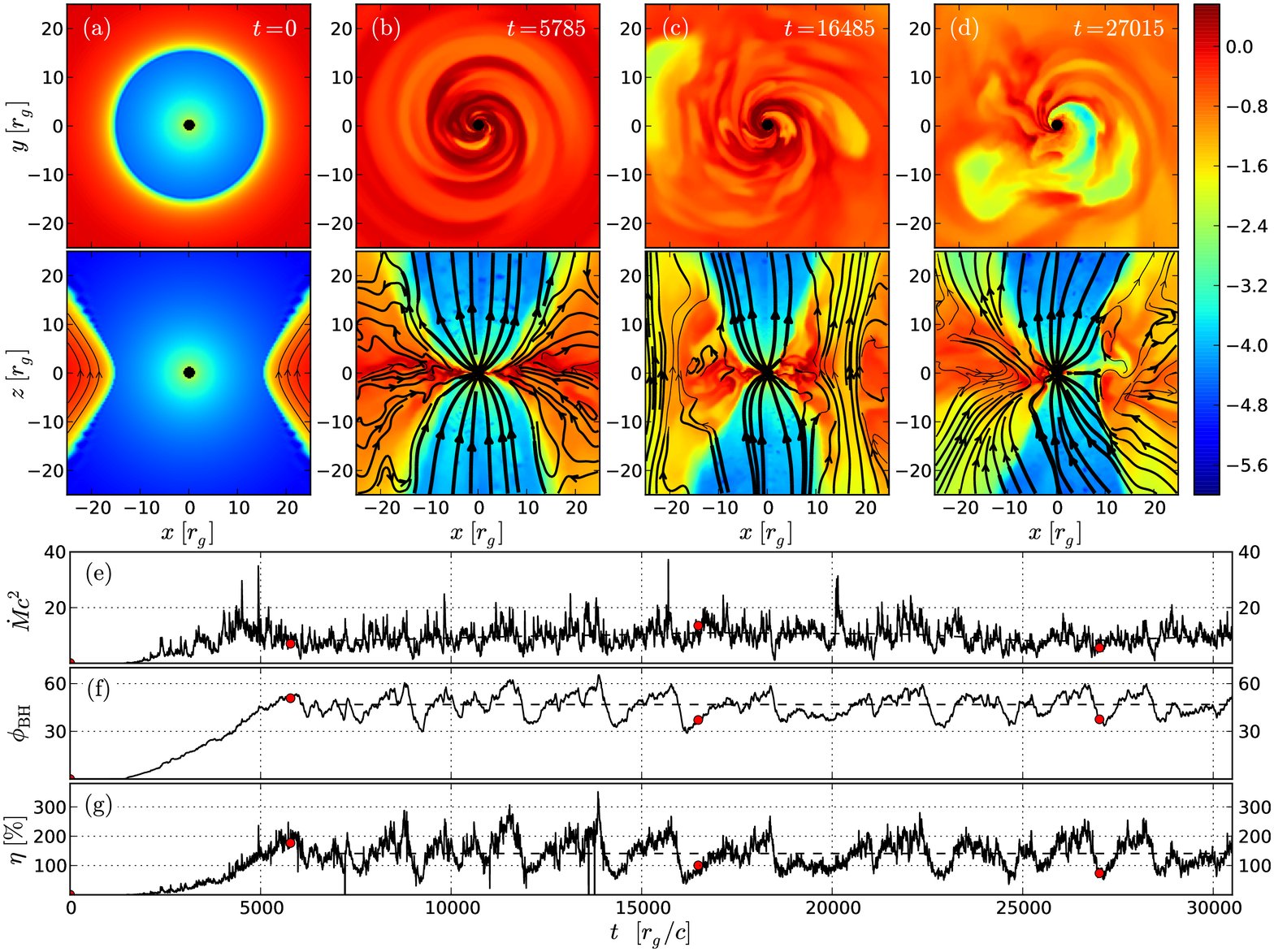}
\caption{Large-scale vertical magnetic field accumulates at the center and forms a dynamically important magnetic field that obstructs the accretion flow and leads to a magnetically arrested disk \citep{2011MNRAS.418L..79T}. Color shows the logarithm of density (see color bar). (a)~The horizontal and vertical slices through the initial condition, a radially extended (from $r_{\rm in}=15r_g$ out to $r_{\rm out} \sim 2\times 10^4r_g$) equilibrium hydrodynamic torus embedded with a weak (plasma $\beta\gtrsim100$) poloidal magnetic flux loop, shown with directed black lines. Even though the magnetic field is not important dynamically, the large size of the torus allows it to hold a large amount of poloidal magnetic flux. (b)~Gas brings in poloidal magnetic flux, which accumulates around the black hole and squeezes the accretion disk vertically. (c,d)~The black hole periodically ejects excess magnetic flux into the disk in the form of low-density magnetic flux eruptions. (e)~Mass-energy accretion rate on the black hole versus time. (f)~Whereas initially the magnetic flux on the black hole monotonically grows in time, at $t\simeq6000r_g$, it becomes dynamically important and partially leaves the black hole, leading to a drop in $\phi_{\rm BH}$. Soon thereafter, the system settles into a quasi steady state: The black hole overeats magnetic flux before shedding its excess into the accretion disk, after which the cycle repeats again. (g)~Jet energy efficiency undergoes oscillations that mirror those of $\phi_{\rm BH}$. The time-average value of the efficiency shown with the dashed line, $\eta\simeq 140\%$, exceeds $100\%$. This is the first demonstration of net energy extraction from an accreting black hole in a numerical simulation.}
\label{fig:mad}
\end{figure}

Numerical simulations are a powerful tool for quantifying the power of relativistic jets and understanding the disk-jet connection. A particularly useful way of doing so is through measuring jet energy efficiency, or the ratio of jet-to-accretion power, $\eta = P_{\rm jet}/\dot Mc^2$. Numerical simulations by different groups found vastly different values of $\eta$: For instance, for $a = 0.99$, the efficiency ranged from $\sim3\%$ \citep{2005ApJ...630L...5M} to $\sim15$\% \citep{2006ApJ...641..103H}. Perhaps not surprisingly, the simulations sample the large parameter space of allowed jet powers, which can range from zero power (no jet) to some maximum power. In the context of magnetically powered jets, this would map into a range from zero large-scale magnetic flux to some maximum value that could be measured by the simulations (see eq.~\ref{eq:pjet}). To determine the maximum jet power, \citet{2011MNRAS.418L..79T} initialized GRMHD simulations with what was then considered an unusually large gas reservoir threaded with vertical magnetic flux (Figure~\ref{fig:mad}a): The initial gas torus size of $20,000r_g$ was much larger than the typical tori of $\sim50r_g$. The large torus size translated into a great amount of vertical magnetic flux contained in the system, thereby flooding the black hole with the magnetic flux and determining the maximum jet efficiency achievable for a given spin. As the simulation started, the MRI led to the accumulation of gas and magnetic flux on the black hole (Figure~\ref{fig:mad}b). This resulted in black hole magnetic flux increasing (Figure~\ref{fig:mad}f) to the point that it overcame the force of gravity that keeps the gas on an orbit around the black hole: The flux became strong enough to escape from the black hole by tearing through the disk (Figure~\ref{fig:mad}c,d). Such periods of flux expulsion followed by periods of relatively quiet accretion and flux accumulation lead to oscillations in both magnetic flux and jet efficiency (Figure~\ref{fig:mad}f,g). In this MAD accretion regime, first simulated in the non-relativistic context  \citep{2003ApJ...592.1042I,2008ApJ...677..317I}, the magnetic flux on the black hole is as strong as possible and is dynamically important.  Its time-average value, $\langle \phi_{\rm BH}\rangle\approx 50$, translates due to eq.~(\ref{eq:pjet}) into the time-average jet efficiency, $\langle \eta \rangle \approx 140\%$. That $\langle\eta\rangle>100\%$ implies that the black hole releases more energy in the form of jets than the entire energy it receives from the accretion flow. Of course, the total energy is conserved, and the extra $40\%$ of energy comes from the black hole spin: Black hole rotation slows down woing to the action of jets. In fact, because of this, in the MAD state black holes slow down to essentially zero spin \citep{2015ASSL..414...45T}.

So far, we have discussed jets from geometrically thick disks with aspect ratio $h/r\sim0.3{-1}$. Radiatively efficient disks, such as those thought to power luminous quasars, are much thinner, with $h/r\sim0.01 (L/0.1L_{\rm E})$.  Unexpectedly, such thin disks led to the production of jets with efficiency $\eta=20{-}50$\% \citep{2019MNRAS.tmp..813L}, defying the theoretical expectations that geometrically thin disks are incapable of dragging large-scale poloidal magnetic flux inward and powering the jets \citep{1994MNRAS.267..235L}. In these simulations, the thin disk was formed by rapid cooling of a thick disk, which is similar to the physics expected to take place during the hard-to-soft spectral state transition. Thus, these powerful jets might represent the transient jets seen in XRBs. They could also be related to radio-loud quasars that make up $10\%$ of all quasars. It is possible that radio quasar is a transient stage in the life of a quasar, and the radio quasars are undergoing a spectral state transition from low-luminosity AGNs to radio-loud quasars. The simulations show signs that these jets are indeed transient phenomena:Tthe magnetic flux strength on the black hole appears to undergo steady decline, indicating that the large-scale magnetic flux diffuses out of the black hole and out to larger radii through the disk, suggesting an impending shutoff of the jets. However, longer-duration simulations are needed to verify this hypothesis.

\subsection{Origin of Large-Scale Vertical Magnetic Flux}
\label{sec:origin-large-scale}

An important question arises: Do black holes in nature receive enough large-scale magnetic flux to produce such powerful jets? Indications are that such strong magnetic fluxes and powerful jets are present in a wide range of systems ranging from AGNs \citep{2014Natur.510..126Z,2014Natur.515..376G,2015MNRAS.449..316N} to tidal disruption events \citep[TDEs;][]{2014MNRAS.437.2744T}\begin{marginnote}[-2pt]\entry{Tidal disruption event (TDE)}{transient emission from the disruption and partial accretion of a star that wanders close to a black hole}\end{marginnote} and gamma-ray bursts \citep{2015MNRAS.447..327T}. What is the origin of this magnetic flux? Whereas in AGNs the ISM contains a substantial amount of large-scale poloidal magnetic flux, sufficient to flood the black hole if the accretion disk can drag the flux toward the black hole \citep{2003PASJ...55L..69N}, the origin of the flux is less clear in TDEs when an unlucky star wanders too close to a supermassive black hole and gets spaghettified by its tidal forces. This is thought to result in an accretion disk that feeds the black hole for months to years. In one remarkable TDE, the formation of the putative disk was accompanied by the launching of a powerful jet \citep{2011Sci...333..203B,2011Natur.476..421B,2011Sci...333..199L,2011Natur.476..425Z} that would require orders-of-magnitude-more large-scale flux on the black hole than that available in the disrupted star \citep{2014MNRAS.437.2744T}. One possibility is that the flux could be provided by a preexisting accretion disk \citep{2014MNRAS.445.3919K}. However, could the large-scale magnetic flux be produced in situ, by the turbulent dynamo in the accretion disk? The problem of large-scale poloidal flux dynamo is long-standing and interdisciplinary: The causes the of 11-year cycle of magnetic flux polarity flips on the Sun and the much longer dynamo cycle in Earth's core \citep{1981ARA&A..19..115C} are still unclear. Numerical simulations indicate that the formation of powerful jets requires preexisting large-scale vertical magnetic flux in the flow \citep{2008ApJ...678.1180B,2012MNRAS.423.3083M}. More recently, evidence points to the large-scale poloidal flux dynamo producing large-scale magnetic flux at least in some circumstances; in fact, in thick disks, $h/r\sim 0.3$, the generated magnetic flux becomes so strong that it leads to the formation of a MAD, even in the absence of any vertical magnetic flux to start with \citep{2020MNRAS.494.3656L}. However, thinner disks $h/r\sim 0.02$, so far do not show evidence of this process \citep{2019arXiv191210192L}.

\subsection{Jet Acceleration}
\label{sec:jet-acceleration}

Many AGN jets reach relativistic velocities, with typical values of bulk Lorentz factor of $\gamma\sim10$. Relativistic motion is not limited to AGNs, but is also observed in XRBs, with $\gamma\sim{\rm few}$, and gamma-ray bursts, with $\gamma\gtrsim100$. How do jets accelerate to such remarkably high velocities? They do so by converting their internal (magnetic, thermal) energy into bulk kinetic energy \citep{1992ApJ...394..459L,2006MNRAS.367..375B}. Typically, such acceleration is accompanied by collimation. For instance, in the case of a well-studied jet in the M87 galaxy, the Lorentz factor is observed to increase as the jet half-opening angle decreases as a power law in radius, $\theta\propto r^{-0.42\pm0.02}$, over more than five orders of magnitude in distance before levelling off at $\theta \simeq 0.01\ \mathrm{rad} \sim 0.6$ degrees \citep{2013ApJ...775..118N}, and the jet reaches Lorentz factors of $\gamma\sim6$ \citep{2016A&A...595A..54M}. In idealized settings, the jet Lorentz factor increases inversely proportional to its half-opening angle, such that $\gamma \sim \theta^{-1}$ \citep{2008MNRAS.388..551T}. The physics behind this acceleration law is simple: The jet accelerates such that its opening angle $\theta$ does not exceed its beaming angle, $\gamma^{-1}$, so the jet maintains transverse causal contact, so that the interior of the jet avoids running into the edge of the jet. However, this acceleration cannot last indefinitely: Eventually, the jet converts most of its internal energy into bulk motion kinetic energy, and the acceleration levels off at $\gamma\sim\gamma_{\rm max}$. Recently, \citet{2019MNRAS.490.2200C} simulated the propagation of black hole jets over five orders of magnitude in distance and showed that the development of the single-power law jet shape extending for as many orders of magnitude occurs naturally if the jets are collimated by an extended accretion flow. Furthermore, the jet accelerates in a way similar to observations.

\subsection{Jet Emission}
\label{sec:jet-emission}

It is broadly agreed upon that jet emission is due to a combination of synchrotron and inverse Compton processes. However, presently, there is no agreement on what accelerates the emitting electrons. In particular, we do not understand what accelerates the electrons that produce the radio emission that we use as an indicator of jet activity. The problem is even more severe in the origin of high-energy gamma-ray (GeV and TeV) emission, where it is unclear whether the emission is coming from near the event horizon or from large distances. There are several candidate processes potentially responsible for accelerating the emitting electrons and, possibly, positrons. They can be accelerated in magnetospheric gaps near the black hole event horizon \citep[e.g.,][]{1998ApJ...497..563H,2015ApJ...809...97B,2016ApJ...818...50H,2016A&A...593A...8P,2017PhRvD..96l3006L,2018ApJ...863L..31C,Parfrey2019}, in 3D magnetic kink instabilities and shocks that can be triggered by jets running into the ISM \citep[e.g.,][]{2016MNRAS.456.1739B,2016MNRAS.461L..46T,2017MNRAS.469.4957B}, and in interface instabilities \citep[e.g.,][]{2019MNRAS.490.2200C}. Heating due to magnetized turbulence in the exterior to the jet is also possible, so what we are seeing could be the emission from the sheath that surrounds the jet (and not from the jet itself \citealt[e.g.,][]{2013A&A...559L...3M,2017MNRAS.467.3604R}).

\subsection{Tilted Disk Precession, Jets, Alignment, and Tearing}
\label{sec:disk-tearing}

A standard approach to modeling black hole accretion is to consider an accretion disk lying in the equatorial plane of the black hole. However, typically the disk midplane is tilted relative to that of the black hole, necessitating the consideration of tilted disks. Early simulations of tilted thick disks with $h/r\sim 0.2$ \citep{2005ApJ...623..347F,2007ApJ...668..417F} confirmed the analytical expectations that general relativistic frame dragging by spinning black holes causes the tilted disks to undergo solid-body precession. However, what happens to their jets? Do they point along the black hole spin axis or disk rotation axis? The jets fly out along the direction of the disk rotation axis \citep{2013Sci...339...49M,2018MNRAS.474L..81L}. However, if the magnetic field is dynamically important, the black hole manages to bring the disk and the jet into alignment at small radii, with the jet initially flying out along the black hole rotational axis before aligning with the rotational axis of the disk at large radii \citep{2013Sci...339...49M}. As the disk precesses, the large-scale jet precesses as well \citep{2018MNRAS.474L..81L,2019arXiv190105970L}, enabling the use of precessing jets as probes of strong-field gravity and general relativistic frame dragging.

The response to the tilt of thinner disks is qualitatively different than it is to that of thick disks. In fact, when the disk thickness is smaller than the disk viscosity parameter, $h/r<\alpha$, warps propagate in the disks viscously.\begin{marginnote}[-2pt]\entry{Bardeen-Petterson effect (BP)}{an abrupt change in the tilt of the disk due to the interaction of viscosity with relativistic precession}\end{marginnote}  \citet{1975ApJ...195L..65B} predicted that this would lead to the inner parts of the accretion disk aligning with the black hole midplane and separating from the outer, misaligned part of the disk by a smooth warp (the Bardeen-Petterson effect or BP). However, at the time it was impossible to take into account the magnetized turbulence responsible for the accretion.\begin{marginnote}[-2pt]\entry{Smoothed particle hydrodynamics (SPH)}{a method for solving fluid dynamics by following the trajectories of particles rather than integrating equations on a mesh}\end{marginnote} And, though such alignment was seen in non-elativistic smoothed particle hydrodynamics (SPH) simulations \citep[e.g.,][]{2000MNRAS.315..570N}, until recently general relativistic magnetized simulations of tilted disks, as thin as $h/r=0.08$, have shown no sign of the BP alignment \citep{2014ApJ...796..104Z}. \citet{2019MNRAS.tmp..813L} simulated the thinnest disk, with $h/r = 0.03$, tilted by $10$ degrees relative to a rapidly spinning black hole, and found that the inner regions of the disk, $r\lesssim5r_g$ aligned with the black hole. This remarkable discovery is the first demonstration of BP alignment in a general relativistic numerical simulation of a magnetized accretion disk. Importantly, the duration of these simulations is long enough for the inner regions of the accretion disk to achieve inflow equilibrium and quasi-steady state. The BP alignment was also seen in shorter nonrelativistic simulations that approximated the effects of black hole rotation via additional torques in MHD equations \citep{2019ApJ...878..149H} and did not include the effects of apsidal precession, which can qualitatively affect the alignment process \citep{2016MNRAS.455L..62N}.

\begin{figure}[h]
\includegraphics[width=0.5\textwidth]{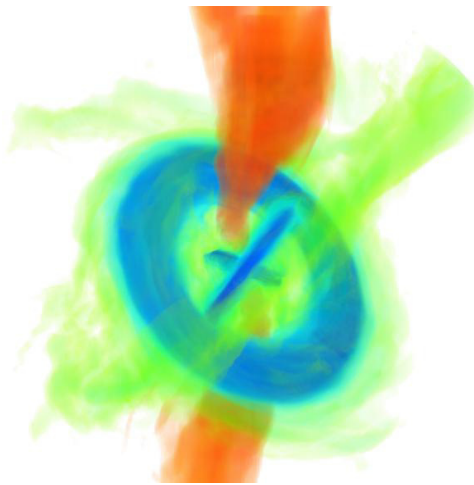}
\caption{Tilted thin accretion disks tear up into individual, independently precessing subdisks.  Whereas at small radii the jets align with the inner subdisks, at large radii the jets align with the outer subdisk. As the disks precess, jets run into subdisks that get in the way and can not only expel them, as seen in the movie of the simulation ({https://youtu.be/mbnG5\_UTTdk}), but also lead to energy dissipation that modifies the radial emission profile of the disk. Figure adapated from \citet{2020MNRAS.tmp..707L}.}
\label{fig:tearing}
\end{figure}

Interestingly, a disk tilted by $65$ degrees got torn up into several individually precessing subdisks \citep{2020MNRAS.tmp..707L}, as seen in Fig.~\ref{fig:tearing}. Note that tilted disks in SPH simulations also show tearing but with an important difference: Tilted disks get torn into a large number of thin rings \citep{2012ApJ...757L..24N}. This difference likely stems from the effects of magnetized turbulence and large-scale magnetic fields that hold the disk together differently than hydrodynamic viscosity. The observational manifestations of torn disks are far reaching. The interactions between adjacent disks can lead to additional dissipation and emission at large radii. Jets and radiation from the inner disk interacting with outer subdisks can contribute additional heating. Not all gas flows between the adjacent subdisks, and an interesting fraction blown away owing to the interaction with the jets and winds. These factors can lead to modifications in the emission profile, potentially reducing the tension with the observed disk sizes (see Fig.~\ref{f:sizes}). Disk tearing can also lead to a wide range of variability, for example, sub-disks precessing through our line of sight might act as absorbers, from time to time dimming the emission of the central source.

\subsection{Jet Interaction with Ambient Medium}
\label{sec:jet-feedback}

\begin{figure}[t]
  \begin{center}
\includegraphics[width=4in,trim=1cm 3cm 1cm 3cm,clip]{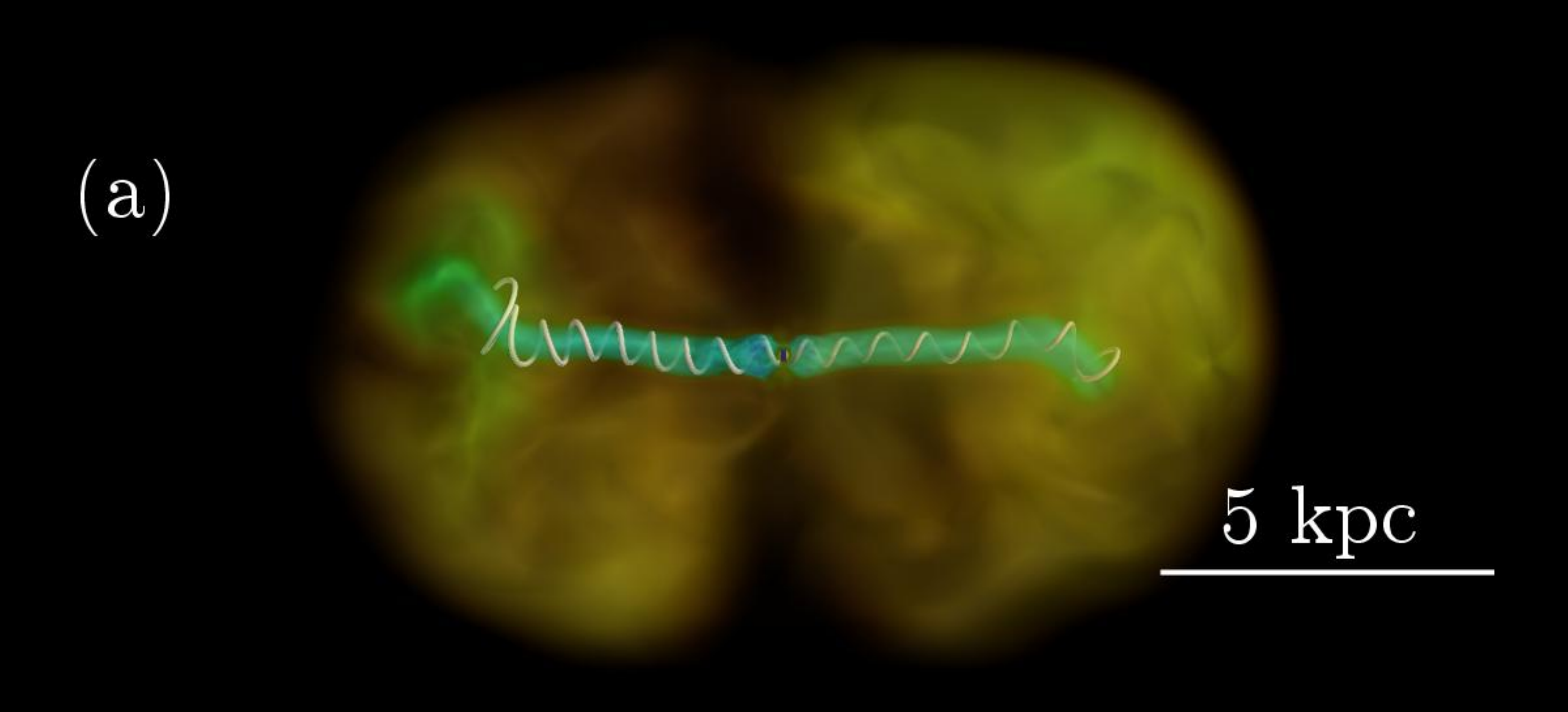}
\includegraphics[width=4in,trim=1cm 3cm 1cm 1.6cm,clip]{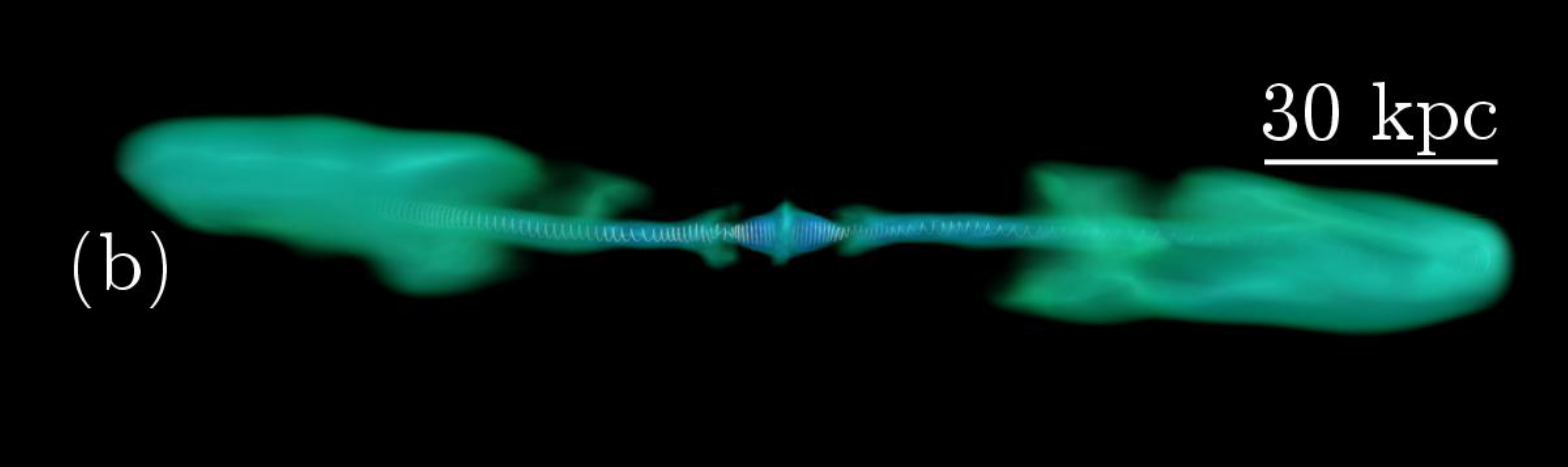}
\end{center}
\caption{
  (a)~Low-power AGN jets (blue-green) succumb
  to global magnetic instabilities, stall within their host galaxies,
  and inflate quasi-spherical cavities (yellow). (b)~High-power jets maintain their stability, leave their host
  galaxies, and form strong backflows. Thus, magnetic instabilities
  can be the key to resolving a 40-year long puzzle on the cause of
    \citet{1974MNRAS.167P..31F} morphological dichotomy of AGNs \citep{2016MNRAS.461L..46T}.}
\label{fig:3d_agn_jets}
\end{figure}

As the jets emerge from the black hole's sphere of influence, they run into the ISM. At this point, their behavior qualitatively changes. In the case of the M87 jet, the jet stops collimating (Section~\ref{sec:jet-acceleration}) and starts to decelerate \citep{2016A&A...595A..54M}. Jet interaction with the ISM is poorly understood. For instance, there is no agreement on the origin of the \citet{1974MNRAS.167P..31F} morphological dichotomy of AGN jets: Fanaroff-Riley type I (FRI) jets appear to develop instabilities early on and often disrupt inside the galaxy, whereas FRII jets appear well collimated and stably propagate to outside of the galaxy. This can have important consequences for their parent galaxy: Whereas FRII jets leave the galaxy unscathed and deposit their energy outside the galaxy, FRI jets inject their energy into the galaxy and can significantly affect the star formation and dynamics of gas. Thus, it is important to understand the stability properties of the jets.  Whereas it is rather easy to reproduce the stable FRII jet morphology in numerical hydrodynamic simulations \citep{1986ApJ...311L..63C}, reproducing the FRI morphology turned out to be much more difficult. Some of the possibilities include
Kelvin-Helmholtz (KH) instabilities in the shear layers %
(e.g.,
\citealt{1997MNRAS.286..215K}; \citealt*{2005A&A...443..863P};
\citealt{2009ApJ...705.1594M}; \citealt{2010A&A...519A..41P}),
and mass entrainment from stellar winds
\citep{1994MNRAS.269..394K,2014MNRAS.441.1488P,wykes_internal_2015}.
 Studies of magnetized jet stability were more promising; however, it turns out that magnetized jet stability sensitively depends on the degree of azimuthal winding of the magnetic field \citep[e.g.,][]{2014ApJ...781...48G}, a free parameter whose value is poorly understood. This introduces an uncertainty into the factors that control jet stability. \citet{2016MNRAS.461L..46T} carried out large-scale simulations of magnetized relativistic jets interacting with the ISM that for the first time were launched via the rotation at the base, as they are launched in nature. The advantage of this approach is that it organically determines the degree of azimuthal winding and therefore leads to the same stability properties of the jets as in nature. Figure~\ref{fig:3d_agn_jets} shows that a change by two orders of magnitude in jet power leads to a drastic change in jet morphology: More powerful jets are stable, whereas less powerful jets are unstable, and this is in qualitative agreement with the FRI/II dichotomy. Future work including magnetic fields and jet precession, which naturally emerges as a result of tilted accretion, will help to refine the models of jet feedback so that they can be used as subgrid models in cosmological simulations.

\section{SUMMARY AND OUTLOOK}

Our understanding of the central engines of AGNs has greatly advanced over past several decades.  We have a fiducial model for how accretion proceeds in the general relativistic spacetime of a supermassive black hole and gained an understanding of the central role played by magnetic fields in the process of jet formation and angular momentum transport.  This general picture of accretion-powered emission is well supported by the existing observational constraints and we owe many of these insights (in part) to the application of state-of-the-art numerical simulations of GRMHD flows.  This is particularly true for our understanding of jet production and jet-disk interactions.  We are beginning to understand the effects of magnetic field geometry on the dynamics of accretion flows, the ability of disks to transport large-scale magnetic fields, and the disk-jet connection. The jets appear to be more resilient than previously thought and defy the standard expectations: They emerge in systems (a) without any large-scale vertical magnetic flux, (b) with a large disk tilt, and (c) with extremely small disk thickness.

However, many challenges remain with our theoretical understanding of accretion disks. Longstanding theoretical inconsistencies remain unresolved, and detailed comparisons between observations and theory yield significant discrepancies.  What many viewed as promising early agreement between theory and observations has faltered in the face of new observational constraints: far-UV SEDs, size-scale constraints, and variability studies.  For these reasons, we have chosen to focus much of this review on the limitations of our current understanding, with the hope of providing a framework to motivate future research.  We have argued that global numerical simulations will be essential for solving some of the most important theoretical issues, including the following:

\begin{itemize}
\item
It is possible that gas fed to the accretion flows on large scales may be misaligned with the black hole spin. We must understand the warping of the accretion flow in such cases, including the possible impact of disk tearing, and its impact on disk emission and reprocessing.

\item
Standard accretion disk models predict the radiation pressure-dominated central regions of accretion flows in AGNs to be violently unstable to thermal and inflow instabilities, yet there is no clear evidence of instability in the generic variability of these sources.  Local MHD simulations cannot address inflow instability and provide conflicting results on the presence of thermal instability. 

\item
Local simulations and simple models suggest that opacities due to atomic transitions could strongly modify the structure of AGN accretion flows or even drive outflows.  Such dynamics are already seen in simulations of massive stellar envelopes.

\item
 Magnetic pressure support seems to be a promising candidate for stabilizing accretion flows in the radiation pressure-dominated regime.  Such support also leads to thicker accretion flows and both may change the vertical structure of accretion flows and increase the reprocessing of emission from the inner regions of the flow.
 
\end{itemize}
All the above questions require global simulations with large dynamic range.  They may also require resolving relatively thin accretion flows, making them computationally expensive.  Many of these questions also require increasingly complex treatments of radiation transfer built on top of already sophisticated algorithms for evolving GRMHD.  Progress will almost certainly require utilizing efficiently optimized algorithms that can harness the capabilities of the next generations of computing infrastructure (exascale and beyond).  Fortunately, there has already been substantial recent progress in developing the numerical tools and algorithms needed for the future, and we are optimistic these questions can be addressed in the next decade.

\begin{issues}[FUTURE ISSUES]
\begin{enumerate}
\item Although local simulations have advantages for studying accretion disks at high resolution, questions about convergence and dependence on simulation domain sizes and magnetically powered outflows strongly motivate global numerical simulations of accretion flows. The large dynamic range required by such simulations requires the  development of algorithms and codes that model the required physics while efficiently utilizing the latest advances in high-performance computing infrastructure.
\item Multiscale simulations that bridge the scales of the galaxy and that of the black hole will be needed to self-consistently determine the mass supply near the black hole and the feedback of the black hole on the galaxy.
\item Realistic treatments of radiation transfer are essential for modeling the radiation-dominated regions of accretion flows to resolve longstanding questions about stability and to study the effects of atomic opacities. The ability of scaling simulations to different black hole masses that is present in most previous nonradiative simulations is no longer possible when radiation transfer is included, requiring AGN- and XRB-specific simulations.
\item Future simulations will need to address the evident discrepancies between observations of AGNs and standard disk models.  This includes size discrepancies, flattening of spectra in the UV and absence of edges, the origin of AGN continuum variability, and lack of generic evidence of instability.
\end{enumerate}
\end{issues}

\section*{DISCLOSURE STATEMENT}
The authors are not aware of any affiliations, memberships, funding, or financial holdings that
might be perceived as affecting the objectivity of this review. 

\section*{ACKNOWLEDGMENTS}
We thank the anonymous referees for many helpful suggestions that improved the quality of this review.  We are grateful to our many colleagues and collaborators, too numerous to list here, who have enlightened us over the years and shaped our views on AGN, accretion disks and jets. 

%

\bibliographystyle{ar-style2}
\bibliography{references}

\end{document}